\documentclass[11pt]{article}

\usepackage[english]{babel}
\usepackage[utf8x]{inputenc}
\usepackage[T1]{fontenc}

\usepackage{amstext,amsmath,amssymb,amsfonts,bbm}
\usepackage{authblk} %Support for footnote style author/affiliation
\usepackage{caption}
\usepackage{graphicx}
\usepackage{color}
\usepackage[colorinlistoftodos]{todonotes}
\usepackage[colorlinks=true, allcolors=blue]{hyperref}
\usepackage{physics} %bunch of new commands for physicists (e.g. \Tr )
\usepackage{slashed} %in order to slash derivatives (gamma matrices)
\usepackage{bbold} %for the bold unit element
\usepackage{bm}%for bold elements
\usepackage{dsfont} % for special fonts (e.g. a different 1 than bbm)

\usepackage[lmargin=80pt,rmargin=80pt,tmargin=80pt,bmargin=80pt]{geometry}
\newcommand{\be}{\begin{equation}}
\newcommand{\ee}{\end{equation}}

\newcommand{\f}{\frac}
\newcommand{\p}{\partial}

\newcommand{\la}{\langle}
\newcommand{\ra}{\rangle}
\let\a=\alpha \let\b=\beta  \let\g=\gamma  \let\d=\delta
       \let\k=\kappa \let\l=\lambda
\let\m=\mu              \let\r=\rho 
\let\s=\sigma \let\t=\tau   \let\vph=\varphi  
    
\let\G=\Gamma \let\D=\Delta   \let\L=\Lambda 
         
  \let\eps=\epsilon

\newcommand{\xb}{\bar{x}}
\newcommand{\alphab}{\bar{\alpha}}
\newcommand{\betab}{\bar{\beta}}
\newcommand{\gammab}{\bar{\gamma}}
\newcommand{\deltab}{\bar{\delta}}
\newcommand{\phib}{\bar{\phi}}
\newcommand{\psib}{\bar{\psi}}
\newcommand{\vphb}{\bar{\varphi}}

\newcommand{\lt}{\tilde{\lambda}}

\newcommand{\Mt}{\tilde{M}}

\newcommand{\cN}{\mathcal{N}}

\begin{document}

\title{\bf Phase diagram and fixed points of\\ tensorial Gross-Neveu models in three dimensions}
\author{Dario Benedetti}
\author{Nicolas Delporte}

\affil{\normalsize\it Laboratoire de Physique Th\'eorique (UMR8627), CNRS, Univ.Paris-Sud, \authorcr
\it Universit\'e Paris-Saclay, 91405 Orsay, France \authorcr
email: dario.benedetti@th.u-psud.fr ,  nicolas.delporte@th.u-psud.fr \authorcr \hfill }

\date{}

\maketitle

\hrule\bigskip

\begin{abstract}
Perturbing the standard Gross-Neveu model for $N^3$ fermions by quartic interactions with the appropriate tensorial contraction patterns, we reduce the original $U(N^3)$ symmetry to either $U(N)\times U(N^2)$ or $U(N)\times U(N)\times U(N)$. In the large-$N$ limit, we show that in three dimensions such models admit new ultraviolet fixed points with reduced symmetry, besides the well-known one with maximal symmetry. 
The phase diagram notably presents a new phase with spontaneous symmetry breaking of one $U(N)$ component of the symmetry group.

\end{abstract}
\bigskip
\hrule\bigskip

%\newpage
\tableofcontents

%%%%%%%%%%%%%%%%%%%%%%%%%%%%%
\section{Introduction}
\label{S:1}
%%%%%%%%%%%%%%%%%%%%%%%%%%%%%

Symmetry principles play a fundamental role in quantum field theory, in particular as they tightly constrain the space of possible interactions, and thus they are essential in the determination of universality classes. When dealing with a large number of degrees of freedom the role of symmetries becomes almost indispensable. 
Consider as an example the case of $\cN$ complex scalar fields $\phi_i$, $i=1,\ldots,\cN$: in the absence of a symmetry relating the different fields, the number of possible quartic interactions is of order $\cN^4$; if instead we demand invariance under $U(\cN)$ transformations we remain with only one interaction, namely $(\sum_i \phib_i\phi_i)^2$. 
The reduction is so drastic that it allows one to actually solve the theory in the limit $\cN\to \infty$, where only a particularly simple class of Feynman diagrams survive \cite{Wilson:1972cf,Coleman:1974jh}.
Things get in general more complicated if we take a smaller symmetry group. An important case is obtained if we write $\cN= N^2$ and we explicitly break the symmetry group from $U(N^2)$ to $U(N)^2$; then it is natural to rearrange the fields as a complex $N\times N$ matrix $\phi_{ab}$, transforming in the fundamental of $U(N)^2$. The large-$N$ limit of matrix models has been extensively explored for its connection to two-dimensional quantum gravity \cite{DiFrancesco:1993cyw} and for its role in AdS/CFT \cite{Aharony:1999ti}, but in general it does not lead to solvable models above two dimensions: the class of dominant graphs (the planar graphs) is still too big and it does not lead for example to a closed equation for the two point function.
One would then imagine that things get even tougher if we shrink further the ratio between the dimension of the symmetry group and the number of fields, but it turns out not to be the case. Writing $\cN= N^3$ and breaking the symmetry group from $U(N^3)$ to $U(N)^3$ leads us into the realm of tensor models: the fields are now arranged as a tensor $\phi_{abc}$ transforming in the fundamental of $U(N)^3$, and the large-$N$ limit of such models has been shown to lead to a restricted subset of the planar diagrams, the so-called melonic graphs \cite{Bonzom:2011zz,Bonzom:2012hw}.\footnote{Most commonly matrix models are presented as describing fields transforming in the adjoint representation, e.g.\ of $U(N)$. Such point of view is natural when introducing them as gauge fields, and wishing to discuss the different behavior of fields in the fundamental representation (flavor fields) and in the adjoint (connection fields) of the same group. Here we discuss them instead as fields in the fundamental representation of a smaller group, because we want to highlight the role of the smaller symmetry group for a given set of fields. Furthermore, the existence of a melonic large-$N$ limit for tensors transforming in an irreducible representation of $U(N)$ (or $O(N)$) has been shown only very recently \cite{Klebanov:2017nlk,Benedetti:2017qxl,Carrozza:2018ewt}, and such models are less understood.}
The melonic dominance typically leads to closed Schwinger-Dyson equations, and thus such models are more manageable than matrix models.\footnote{Another way to obtain similar equations is to consider multi-matrix models at large $N$ and in the limit of large number of matrices (see \cite{Azeyanagi:2017mre} and references therein).}

Given such discovery, and knowing that vector and matrix models have a great wealth of applications, it is not unreasonable to imagine that tensor models might hold the key to a panoply of new theoretical insights.
While the zero dimensional models have been explored for some years in connection to their quantum gravity interpretation \cite{GurauRyan-review,Gurau-book,Bonzom:2012wa,Bonzom:2014oua,Carrozza:2015adg,Tanasa:2015uhr,Bonzom:2016dwy}, which was also the reason behind their original introduction \cite{Ambjorn:1990ge,Sasakura:1990fs}, higher dimensional tensor models are a very recent entry in the theoretical landscape. 

A great boost came with the observation \cite{Witten:2016iux} that in one dimension they provide an alternative to the Sachdev-Ye-Kitaev model \cite{Sachdev:1992fk, Kitaev2015, Maldacena:2016hyu, Polchinski:2016xgd} dispensing with the quenched disorder of the latter. As a consequence, quantum mechanical tensor models have been the subject of several studies, see for example \cite{Klebanov:2016xxf,Peng:2016mxj,Krishnan:2016bvg,Krishnan:2017lra,Bulycheva:2017ilt,Choudhury:2017tax,Halmagyi:2017leq,Klebanov:2018nfp,Chang:2018sve,Carrozza:2018psc} (see also \cite{Delporte:2018iyf,Rosenhaus:2018dtp,Klebanov:2018fzb} for reviews). Tensor models in two or more dimensions remain so far the least explored, one reason being that when seeking direct generalizations of the SYK-type Schwinger-Dyson equations (i.e.\ in models with the so-called tetrahedron interaction) to higher dimensions one finds no non-trivial conformal theories in the critical dimension \cite{Benedetti:2017fmp} or conformal theories with a complex spectrum, except in small non-integer ranges of the spacetime dimension \cite{Giombi:2017dtl,Prakash:2017hwq,Giombi:2018qgp}.

In this paper we wish to push further the exploration of new tensorial conformal field theories in higher dimensions, with slightly different motivations.
First, in the $U(N)^3$-symmetric version of the Gross-Neveu model studied in \cite{Benedetti:2017fmp} (which does not allow for a tetrahedron interaction), it was found that one of the $U(N)$ subgroups can be spontaneously broken, which is a genuinely new feature with respect to the usual vector case. However, that model being in two dimensions, such breaking was identified as a large-$N$ artefact (as for continuous chiral symmetry breaking in the usual vector case \cite{Witten:1978qu}), in agreement with the Coleman-Mermin-Wagner theorem \cite{Mermin:1966fe,Coleman:1973ci}.
We would like to check here whether such new broken phase survives in three dimensions, where the Coleman-Mermin-Wagner theorem does not apply.
Second, the vectorial Gross-Neveu model is known to have a non-trivial UV fixed point in three dimensions \cite{Rosenstein-PRL,deCalan:1991km,ROSENSTEINreview}, hence it is an intriguing question whether it remains the only non-trivial fixed point upon a tensorial extension, or new ones are to be found.
And we believe it would be interesting to identify new conformal field theories of tensor type, even if these are not a direct generalization of the SYK type of conformal theory. In particular, given the point of view we sketched above, with the tensor models as symmetry-breaking perturbations of the vector models, it is natural to wonder whether such new critical theories would correspond to deformations of the Klebanov-Polyakov duality \cite{Klebanov:2002ja}. 
The latter is a duality between the singlet sector of the either the free or Wilson-Fisher fixed points of the vector $O(N)$ model in three dimensions and Vasiliev's (type A) higher spin theory in $AdS_4$ with two different boundary conditions. It has been generalized to the case of fermions \cite{Sezgin:2003pt}, thus relating the free or critical Gross-Neveu model in three dimensions to type B higher spin theory in $AdS_4$ with the corresponding boundary conditions (see \cite{Giombi:2016ejx} for a review). Recently, Vasiliev proposed a new type of higher spin theory that could be dual to tensor theories \cite{Vasiliev:2018zer}, therefore the question of whether interacting fixed points exist for tensor models is of interest.

With such motivations in mind, we present here two modifications of the three-dimensional Gross-Neveu model having $U(N)\times U(N^2)$ and $U(N)^3$ symmetry, respectively. The first is in fact rather a rectangular matrix model, but it has many features in common with the second, properly tensorial, model. Studying the models at large $N$ by means of Schwinger-Dyson equations, and computing the effective potential for the intermediate fields, we confirm the presence of a phase with spontaneously broken $U(N)$ subgroup, and we find two new interacting fixed points in each model.

%%%%%%%%%%%%%%%%%%%%%%%%%%%%%
\section{A brief reminder of the vectorial Gross-Neveu model}
\label{sec:vectorGN}
%%%%%%%%%%%%%%%%%%%%%%%%%%%%%
The Gross-Neveu model \cite{Gross:1974jv} has been extensively studied, in particular in two dimensions, where it provides a model of asymptotic freedom and dynamical mass generation, which is also integrable. Here, we are rather interested in its three-dimensional version, which despite being perturbatively non-renormalizable, is renormalizable in the $1/N$ expansion \cite{Parisi:1975im} and admits an ultraviolet fixed point at large $N$ \cite{Rosenstein-PRL,deCalan:1991km} which renders the model meaningful at arbitrarily high energies (see \cite{ROSENSTEINreview} for a review). The nontrivial fixed point theory has been conjectured to be dual to a particular version of higher spin theory in $AdS_4$ \cite{Sezgin:2003pt}, a conjecture which has passed several tests (see \cite{Giombi:2016ejx} and references therein).

In view of the upcoming generalizations, we define here the model for the case of $N^3$ Dirac fermions in Euclidean signature (see Appendix \ref{sec:appendix gamma} for conventions on $\g$ matrices). The action is\footnote{Here and in the rest of the paper repeated indices imply a summation.}
\be
S_{\rm GN}[\psi,\bar{\psi}] = \int d^3 x \; \left( \psib_i  \slashed{\p} \psi_i -\f{\l}{N^3} (\psib_i \psi_i)^2 \right) \;.
\ee
Expressing the four-fermion interaction in terms of an intermediate field $\sigma$, the action writes
\begin{equation} \label{eq:GNY}
S[\psi,\bar{\psi},\sigma]=\int d^3 x \;\left(\bar{\psi}_i\slashed{\partial}\psi_i + \sigma\bar{\psi}_i\psi_i + \frac{N^3}{4\l}\sigma^2\right).
\end{equation}
Besides the $U(N^3)$ invariance (with the fermions transforming in the fundamental representation), the model has also a discrete chiral symmetry, which acts as
\begin{equation}
\psi\rightarrow \gamma^5\psi \quad \bar{\psi} \rightarrow -\bar{\psi}\gamma^5 \quad \sigma \rightarrow - \sigma.
\end{equation}

In the large-$N$ limit one can write a closed Schwinger-Dyson equation for the fermion 2-point function, which reduces to a gap equation for the fermion mass $m=\expval{\sigma}$:
\begin{equation} \label{eq:GN-gap}
\frac{m}{\l} = 8 m \int_\Lambda \frac{\dd[3]p}{(2\pi)^3}\frac{1}{p^2+m^2}\;,
\end{equation}
where the divergent integral is regulated by a UV cutoff $\L$. The integral on the right-hand side of the gap equation is a monotonically decreasing function of $m$, hence it has a maximum at $m=0$, which defines a critical coupling
\begin{equation}
\frac{1}{\l_c} \equiv 8\int_\Lambda \frac{\dd[3]p}{(2\pi)^3}\frac{1}{p^2} \;,
\end{equation}
above which the gap equation \eqref{eq:GN-gap} admits a real solution $m\neq 0$, besides the trivial one.
Using the intermediate field formulation \eqref{eq:GNY}, and integrating out the fermions, one finds that for $\l>\l_c$ the stable solution of the effective potential is the non-zero solution. Therefore, the theory has a dynamically generated mass for $\l>\l_c$, and this in turn means that the chiral symmetry is spontaneously broken. 

Using the gap equation for $\l>\l_c$, the effective potential writes
\begin{equation} \label{eq:V_GN}
V_{\rm eff}(\sigma) = \frac{1}{\pi}\left(\frac{1}{3}\abs{\sigma}^3 - \frac{m}{2}\sigma^2\right) \;,
\end{equation}
with an evident minimum at $\s=m$.

For $0\leq \l\leq \l_c$ the symmetry is instead preserved, as $m=0$ is stable. The phase transition at $\l=\l_c$ is second order.

The $\beta$-function of the adimensional coupling $\lt \equiv \L \l$ is obtained from eq.~\eqref{eq:GN-gap} derivating both sides with respect to $\L$, leading to 
\begin{equation}
\b = \L\p_\L \lt = \lt - \frac{4}{\pi^2}\lt^2.
\end{equation}

%%%%%%%%%%%%%%%%%%%%%%%%%%%%%
\section{$U(N)\times U(N^2)$-symmetric model}
\label{sec:S2}
%%%%%%%%%%%%%%%%%%%%%%%%%%%%%

The $U(N)\times U(N^2)$-symmetric model is obtained by first rearranging the label $i=1,\ldots,N^3$ as a set of two labels $a$ and $A$, so that we rewrite $\psi_i \to \psi_{aA}$, with $a=1,\ldots,N$ and $A=1,\ldots,N^2$, and $\psi_{aA}$ transforming in the fundamental of the product group.\footnote{There is a slight redundancy in denoting the symmetry group as $U(N)\times U(N^2)$: its action on $\psi_{aA}$ is not faithful, because the action of the two $U(1)$ subgroups of $U(N)$ and $U(N^2)$ are indistinguishable. Therefore, a faithfully acting symmetry group of the theory would be $U(1)\times (SU(N)/\mathbb{Z}_N\times SU(N^2)/\mathbb{Z}_{N^2})$, where we have quotiented also by the residual centers of the special unitary groups. A similar caveat applies of course also to the symmetry group of section \ref{S:2}. In the rest of the paper, for compactness of notation we will stick to the non-faithful denotation of the symmetry group.} 
In order to explicitly break the symmetry from $U(N^3)$ to $U(N)\times U(N^2)$, while preserving the discrete chiral invariance, we add the following interaction to the GN model:
\be
\frac{\l_p}{N^2} \bar{\psi}_{a A}\psi_{a^\prime A}\bar{\psi}_{a^\prime A'}\psi_{a A'} \;.
\ee

\begin{figure}
\centering
\begin{minipage}{0.4\textwidth}
           \centering 
            \includegraphics[width=0.5\textwidth]{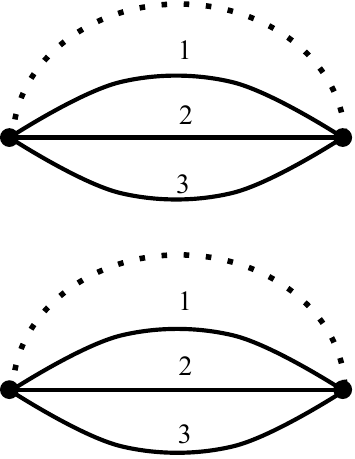}
        \end{minipage}
        \hspace{0.01\textwidth}
\begin{minipage}{0.4\textwidth}
            \centering
            \includegraphics[width=0.5\textwidth]{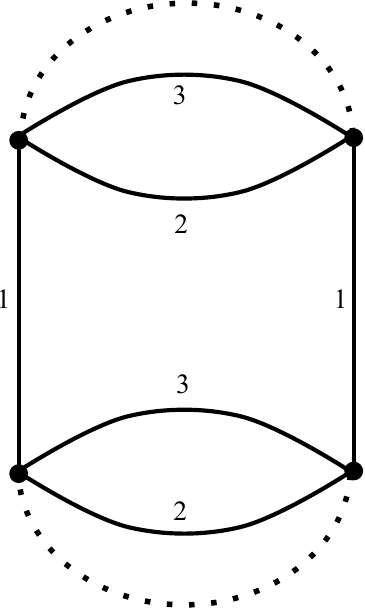}
        \end{minipage}
       % \hspace{0.01\textwidth}
\caption{\label{fig:interactions}Graphical representation of the interaction vertices (GN and pillow).}
\end{figure}

In view of the next generalization, we will actually replace also the index $A$ by a pair of indices, each taking values from 1 to $N$, i.e.\ we write $\psi_i \to \psi_{abc}$. The total action then reads
\be \label{eq:action}
S[\psi,\psib] = S_{\rm free}[\psi,\psib] + S_{\rm int}[\psi,\psib] \;,
\ee
with
\be
S_{\rm free}[\psi,\psib]= \int d^3 x \;  \bar{\psi}_{abc}\slashed{\partial}\psi_{abc} \;,
\ee
\be \label{eq:S_int1}
S_{\rm int}[\psi,\psib]= -\frac{\l}{N^3}  \int d^3 x \; (\bar{\psi}_{abc}\psi_{abc})^2 - \frac{\l_p}{N^2} \int d^3 x \; \bar{\psi}_{abc}\psi_{a^\prime bc}\bar{\psi}_{a^\prime b^\prime c^\prime}\psi_{ab^\prime c^\prime} \;.
\ee
Having written the rectangular matrix as a cubic tensor, we can depict the interactions as in Fig.~\ref{fig:interactions}, where each vertex represents a tensor, and the solid lines with label $n=1,2,3$ represent the contraction of two indices in the $n$-th position. The dotted lines represent instead the spin contraction (as in \cite{Benedetti:2017fmp}, we could consider also other interactions in which such contraction is mediated by a $\g_5$ or $\g_\m$ matrix). The solid-line graph on the right of Fig.~\ref{fig:interactions} is commonly called the \emph{pillow} graph, hence the subscript $p$ for its coupling $\l_p$.
Notice that it comes with a different power of $N$ in \eqref{eq:S_int1}, as required for a non-trivial large-$N$ limit (see for example \cite{Bonzom:2016dwy}).

%%%%%%%%%%%%%%%%%%%%%%%%%%%%%
\subsection{\texorpdfstring{$\beta$}{b}-functions and flow diagram}
\label{sec:SD1}
%%%%%%%%%%%%%%%%%%%%%%%%%%%%%

%%%%%%%%%%%%%%%%
\begin{figure}
\centering 
\includegraphics[width=0.6\textwidth]{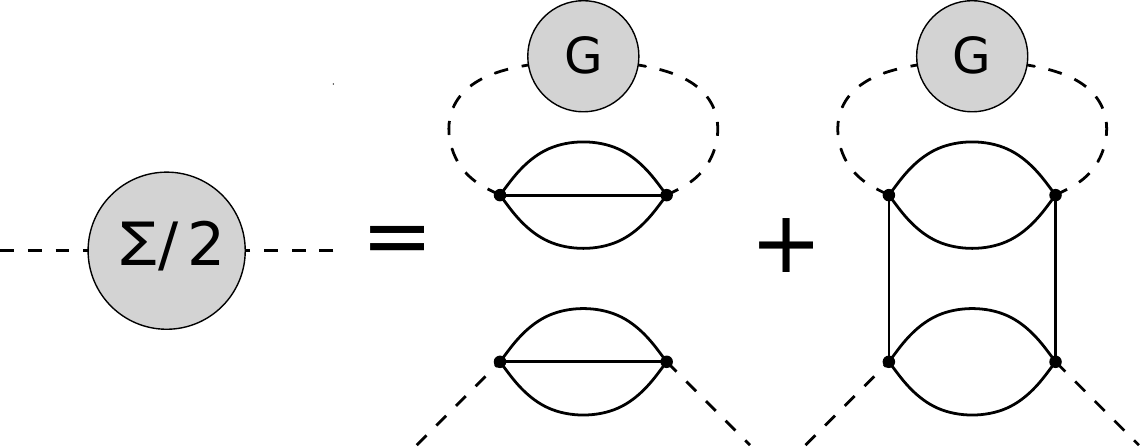}
\caption{\label{fig:SDE}Schwinger-Dyson equation at large $N$ for the self-energy.}
\end{figure}
%%%%%%%%%%%%%%%%

Assuming that the $U(N)\times U(N^2)$ invariance is unbroken, we write the two-point function as
\be \label{eq:2pt-sym-ansatz}
\langle \psi_{a_1 a_2 a_3}(x) \psib_{b_1 b_2 b_3}(x') \rangle = G(x,x')\, \d_{a_1 b_1} \d_{a_2 b_2} \d_{a_3 b_3} \;.
\ee
In the large-$N$ limit, the self-energy $\Sigma$ is expressible in terms of tadpole diagrams with full two-point function on the internal propagator, as depicted on Fig.~\ref{fig:SDE}.
As a consequence, the large-$N$ Schwinger-Dyson equation is a closed equation for $G(x,x')$, which for its Fourier transform $\hat{G}(p)$ reads
\begin{equation}
\hat{G}(p)^{-1} 	= i\slashed{p} -\Sigma(p) = i\slashed{p} + 2(\l + \l_p)\int\frac{\dd[3]q}{(2\pi)^3}\tr[\hat{G}(q)] \mathbb{1}\;,
\end{equation}
the trace and the identity being defined in spinor space (in the following we will generally omit the identity matrix, unless we want to emphasize its presence).
Since the tadpole integral is momentum-independent, we can write $\Sigma = -m\mathbb{1}$, resulting in the gap equation
\begin{equation}
m = 2(\l + \l_p)\int\frac{\dd[3]q}{(2\pi)^3} \frac{\tr(-i\slashed{q} + m)}{q^2 + m^2}, 
\end{equation}
or
\begin{equation}
\frac{1}{\l+\l_p}=8 \int_{\abs{p}<\Lambda}\frac{\dd[3]p}{(2\pi)^3}\frac{1}{p^2+m^2} = \frac{4}{\pi^2}\left(\Lambda - m \arctan\left(\frac{\Lambda}{m}\right)\right)\;,
\label{eq:gap-1c}
\end{equation}
that is the analog of \eqref{eq:GN-gap}.

\begin{figure}[!h]
\begin{minipage}{0.5\textwidth}
  	\centering 
	\includegraphics[width=0.8\textwidth]{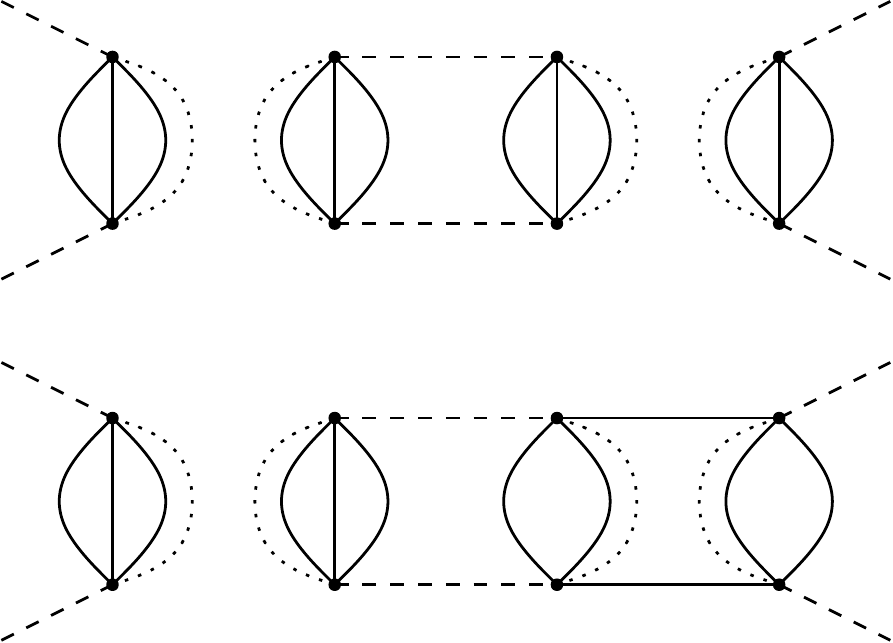}
        \end{minipage}
\begin{minipage}{0.5\textwidth}
      	\centering
	\includegraphics[width=0.8\textwidth]{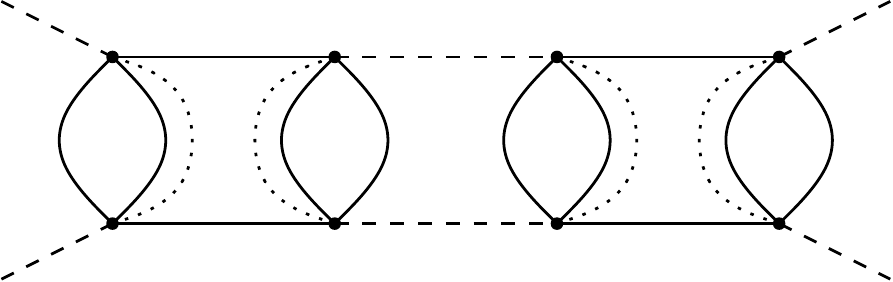}
        \end{minipage}
\caption{\label{fig:4-point graphs}Leading order graphs renormalizing the $\lt$ and $\lt_p$ couplings respectively.}
\end{figure}

\begin{figure}[!h]
\centering
\includegraphics[width=0.6\textwidth]{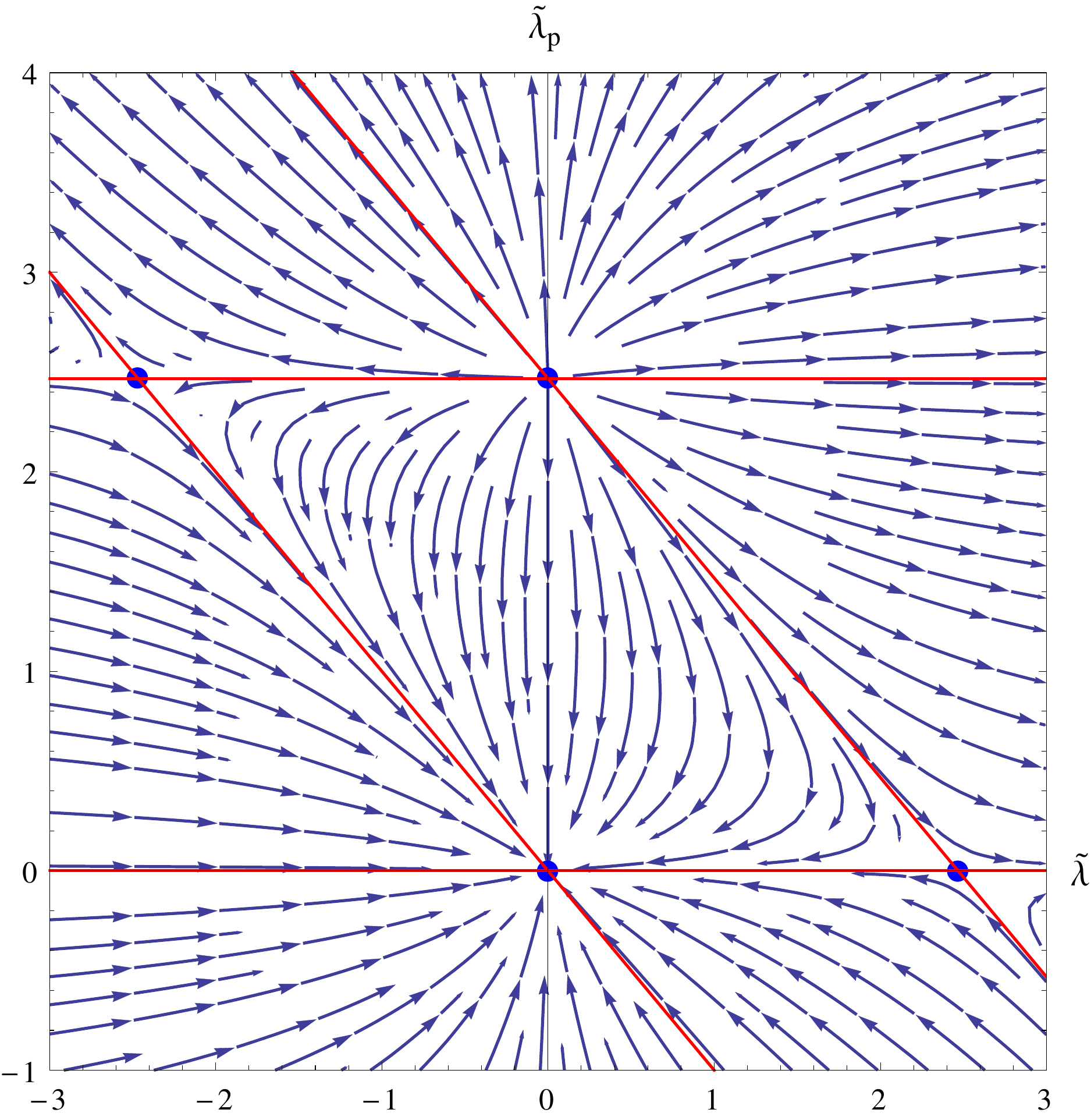}
\caption{Renormalization group flow in the $\left(\lt,\lt_p\right)$-plane. Arrows point towards the $IR$, blue dots denote the fixed points, and red lines mark the zeros of either $\beta_p$ or $\beta+\beta_p$. }
\label{fig:RG-flow-fermions}
\end{figure}

After a rewriting in terms of the dimensionless couplings ($\lt_p \equiv \Lambda\l_p$ and $\lt \equiv \Lambda\l$), and derivating both sides with respect to $\Lambda$, we get 
\begin{equation*}
\frac{1}{\lt+\lt_p} - \frac{\Lambda}{\left(\lt+\lt_p\right)^2}\partial_\Lambda\left(\lt+\lt_p\right) = \frac{4}{\pi^2}\left(1 - \left(\frac{m}{\Lambda}\right)^2\right)\;.
\end{equation*} 
Defining $\kappa \equiv 4/\pi^2$, and taking $\L\gg m$, we find the following combination of beta functions:
\be
\beta+\beta_p  \equiv \Lambda\partial_\Lambda \lt + \Lambda\partial_\Lambda\lt_p= \left(\lt + \lt_p\right) - \kappa\left(\lt + \lt_p\right)^2 \;.
\ee
Taking into account the different structure of diagrams that contribute to the flow of the couplings (see Fig.~\ref{fig:4-point graphs}), we can disentangle the beta functions and obtain
\begin{align}
\beta &= \lt - \kappa\left(\lt^2 + 2\lt\lt_p\right)\;,\\
\beta_p &= \lt_p - \kappa\lt_p^2 \;.
\label{eq:beta-1c}
\end{align}
%%%%%%%%%
%%%%%%%%%
This leads to the flow diagram of Fig.~\ref{fig:RG-flow-fermions}.
%%%%%%%%%
%%%%%%%%%

One sees, in addition to the trivial IR-fixed point $\left(\lt,\lt_p\right) = (0,0)$, three new non-trivial fixed points at $\left(\frac{1}{\kappa},0\right)\equiv {\rm FP}_1$, $\left(-\frac{1}{\kappa},\frac{1}{\kappa}\right)\equiv {\rm FP}_2$, and $\left(0,\frac{1}{\kappa}\right)\equiv {\rm FP}_3$. The first of them is the usual interacting CFT of the vector GN model, while the other two are new interacting CFTs. The fixed point at the origin is IR-stable, while ${\rm FP}_2$ is UV-stable, and the other two are saddles. The irrelevant directions at ${\rm FP}_1$ and ${\rm FP}_3$ can be understood as a statement of the fact that their universality classes are stable against symmetry breaking perturbations (for ${\rm FP}_1$, which being the usual GN fixed point has a larger symmetry, namely $U(N^3)$) and trace perturbations (for ${\rm FP}_3$, which lies on the tracelessness constraint subspace of the diagram).
The critical exponents are all $\pm1$ with the signs determined by the corresponding eigenperturbation being relevant or irrelevant.\footnote{In our convention, the critical exponent corresponds to the mass-dimension of the integrated operator (opposite to that of the relative coupling), hence relevant ones have negative critical exponents.}

%%%%%%%%%%%%%%%%%%%%%%%%%%%%%
\subsection{Effective potential and phase diagram}
\label{subsection: effective potential}
%%%%%%%%%%%%%%%%%%%%%%%%%%%%%
The next question to raise concerns the nature of the phase diagram of the stable vacuum.

Following \cite{Benedetti:2017fmp}, we introduce a Hermitian matrix $M_{ij}$ as intermediate field, such that the interaction terms rewrite as\footnote{In App.~\ref{app:rectangular}, we will explain what happens if instead we choose to introduce an intermediate field by cutting the pillow interaction along the index of size $N^2$ (or double index in the tensor notation).}
\begin{equation}
\label{eq: sint with psi}
S_{\rm int}[\psi,\psib,M]=\frac{1}{2}\left[\Tr(M^2) + \frac{b}{(1-b)N}(\Tr M)^2\right] + \frac{\sqrt{2\l_p}}{N}\bar{\psi}_{ibc}\psi_{jbc}M_{ij},
\end{equation}
$b \equiv \frac{-\l}{\l_p}$, allowing to integrate out the fermions and obtain
\begin{equation}
\label{eq:S_M}
S[M]=\int \frac{1}{2}\left[\Tr(M^2) + \frac{b}{(1-b)N}(\Tr M)^2\right] - N^2 \tr \Tr\left[\ln\left(\slashed{\partial} + \frac{\sqrt{2\l_p}}{N} M\right)\right].
\end{equation}
To remove the $N$ factor from inside of the log-term, we rescale $M \rightarrow NM$. We are interested in the effective potential, which in the large-$N$ limit is simply given by \eqref{eq:S_M} evaluated at constant $M$.\footnote{Remember that the effective potential is defined as the effective action $\G[M]$ (the one-particle-irreducible generating functional) at constant field, and that $\G[M]$ is the Legendre transform of $W[J]$, the generating functional of connected $n$-point functions. Since the latter is given in the large-$N$ limit simply by the Legendre transform of $S[M]$, and since the Legendre transform is an involutive transformation, we conclude that $\G[M]=S[M]$.} Then the last term, up to a constant independent of $M$, gives
\be
\begin{split}
\tr \Tr \int \frac{\dd[3]k}{(2\pi)^3} \ln\left(\slashed{\partial} + \sqrt{2\l_p} M\right) &= -4 \Tr \int \frac{\dd[3]k}{(2\pi)^3}\sum_{n>0}\left(\frac{ik}{k^2}\sqrt{2\l_p} M\right)^{2n}\frac{1}{2n} \\
&= \frac{1}{\pi^2}\int_0^\Lambda \dd k k^2 \Tr\log\left(1 + \frac{2\l_p M^2}{k^2}\right) \\
&= \frac{\Tr}{3\pi^2} \left[4\l_p\Lambda M^2 - 2(2\l_p M^2)^\frac{3}{2}\left(\arctan \frac{\Lambda}{\sqrt{2\l_pM^2}}\right) \right. \\
&\quad\qquad \left. + \Lambda^3\log\left(1 + \frac{2\l_pM^2}{\Lambda^2}\right)\right].
\end{split} 
\ee

Switching to dimensionless variables and couplings, we find the following effective potential:
\begin{equation} \label{eq:effective action}
\begin{split}
V_{\rm eff}[\Mt]\equiv \frac{S[\Mt = \text{const.}]}{N^2\Lambda^3 {\rm Vol}}= & \frac{1}{4\lt_p}\left[\Tr(\Mt^2) + \frac{b}{(1-b)N}(\Tr \Mt)^2\right] \\
& - \frac{1}{3\pi^2}\Tr\left[2 \Mt^2 - 2\abs{\Mt}^3\arctan{\frac{1}{\abs{\Mt}}} + \log\left(1 + \Mt^2\right)\right] \;,
\end{split}
\end{equation}
where we defined
\begin{equation}
\Mt \equiv \frac{\sqrt{2\l_p} M }{\Lambda}, \quad \lt \equiv \l \Lambda, \quad \lt_p \equiv \l_p \Lambda, \quad {\rm Vol}=\int d^3 x \;.
\end{equation}
Owing to the $U(N)$-invariant form of the effective potential \eqref{eq:effective action}, we can diagonalize the matrix $\Mt$ and recover an effective potential for its set of eigenvalues\footnote{The Vandermonde determinant originating in the change of variables is subleading in $1/N$ (the action is of order $N^3$ and the logarithm of the Vandermonde determinant is of order $N^2$, see \cite{Nguyen:2014mga}), hence it is not included.} 
$\mu_i$, $1\leq i\leq N$:
\begin{align} \label{eq:effective potential}
V_{\rm eff}[\{\mu_i\}]  &= \sum_i \left[\frac{1}{4\lt_p}\mu_i^2 + \frac{1}{3\pi^2}\kappa(\mu_i)\right] - \frac{\lt}{4\lt_p (\lt+\lt_p) N}\left(\sum_i\mu_i\right)^2 \;,\\
\kappa(\mu) &= 2\mu^3\arctan\frac{1}{\mu} - 2\mu^2 - \log(1+\mu^2) \;.
\end{align}
An important point is that the potential \eqref{eq:effective potential} is unbounded from below in the regions $\lt_p<0$ and $\lt_p+\lt<0$ (this is most easily seen by studying special symmetric configurations such as those we will encounter below for the stationary points), therefore considered unphysical.

The only extremum of the potential which preserves all the symmetries of \eqref{eq:action} is the trivial solution $M=0$, for which our potential is normalized such that $V_{\rm eff}[0] = 0$.
Stationary points with non-zero eigenvalues $\mu_i = \mu ~ (1\leq i\leq N)$, spontaneously break the chiral symmetry of \eqref{eq:action} (reflected in the symmetry $M \rightarrow -M$), whereas if the eigenvalues are not all equal, the original $U(N)$ symmetry of \eqref{eq:action} is spontaneously broken as well.

\begin{figure}
\centering
\includegraphics[width=0.7\textwidth]{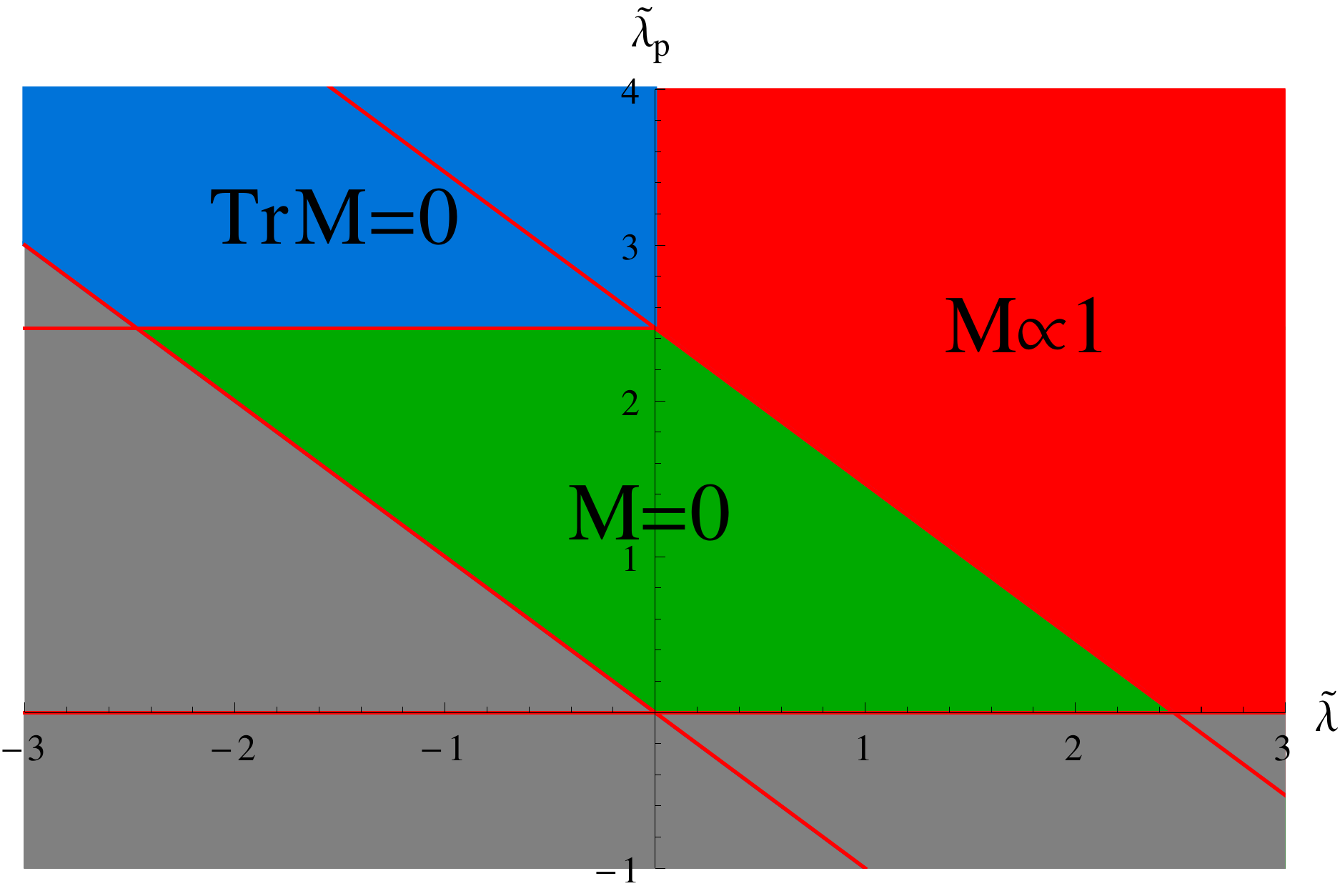}
\caption{\label{fig:phase diagram}Phase diagram of the fermionic TGN model in 3$d$. The red zone has a vacuum $M\propto \mathbb{1}$, the blue one to a traceless vacuum and the green parallelogram to $M=0$. The grey zone corresponds to an unstable model.}
\end{figure}

In the green parallelogram region of the phase diagram in Fig.~\ref{fig:phase diagram}, we can show that the potential is non-negative, and the solution $\mu = 0$ gives a global vacuum. Indeed, the term in square brackets of eq. \eqref{eq:effective potential} is (for each $i$) non-negative and convex (with a global minimum at the origin) in the range $0<\lt_p<\pi^2/4$. 
Consequently, if $\sum_i\mu_i\neq 0$ and $\lt< 0$ then $V_{\rm eff}[\{\mu_i\}]  > 0 = V_{\rm eff}[0]$. In the case $\lt> 0$, we can use the Cauchy-Schwarz inequality to bound
\begin{equation}
V_{\rm eff}[\{\mu_i\}]  \geq 
\sum_i \underbrace{\left[\frac{1}{4\lt_p}\mu_i^2 + \frac{1}{3\pi^2}\kappa(\mu_i) - \frac{\lt}{4\lt_p (\lt+\lt_p)}\mu_i^2\right]}_{\equiv w(\mu_i)},
\end{equation}
which is convenient, as the eigenvalues decouple. By taking first and second derivatives of each term we can now prove that $\mu_i = 0$ is the unique minimum of $w(\mu_i)$, and hence $\mu_i =  0 ~\forall i$ is the global minimum of $V_{\rm eff}[\{\mu_i\}]$ for $0\leq \lt+\lt_p \leq \pi^2/4$ and $0\leq \lt_p \leq \pi^2/4$:
\begin{align}
w^\prime(x)& = \frac{2x}{\pi^2 \alpha}\left[1 - \alpha + \alpha x \arctan\frac{1}{x}\right] >0 \quad (x>0,\; 0<\a<1),\\
w^{\prime\prime}(x) &= \frac{2}{\pi^2 \alpha}\left[1  -\alpha\f{1+2x^2}{1 + x^2}  + 2\alpha x  \arctan\frac{1}{x}\right] >0 \quad (0<\a<1) \;,
\end{align}
having introduced $\alpha= 4(\lt+\lt_p)/\pi^2$. At $\a>1$ the origin becomes unstable.

The stability of the trivial solution is more properly analyzed by studying the full Hessian.
Coming back to eq.~\eqref{eq:effective action}, we can want to compute the second derivative around the point $\Mt = 0$, for which we can discard the $\arctan$ term, as it is of cubic (and higher) order in the fluctuations. The first derivative gives 
\begin{gather}
\pdv{V}{\Mt_{ij}} = A \Mt_{ji} + B \f{\Tr \Mt}{N} \d_{ij}+ C \Mt_{ji} \; ,\\
A = \f{1}{2\lt_p}\; ,\quad B = \f{b}{2\lt_p(1-b)}\; , \quad C = - \f{2}{\pi^2}\;.
\end{gather}
The second derivative
\begin{equation}
\pdv{V}{\Mt_{ij}}{\Mt_{kl}} = (A + C) \d_{ik}\d_{jl} + B\f{\d_{ij}\d_{kl}}{N}\;,
\end{equation}
can be rewritten as follows
\begin{gather}
H = \a(1 - P)+ \b P \\
\a = A + C\; \quad \b = A + B + C\;,\quad P_{ij,kl} \equiv \frac{\d_{ij}\d_{kl}}{N}\;,
\end{gather}
introducing $P$ that projects on the trace.

Articulated as such, the Hessian $H$ is easy to diagonalize, as the eigenfunctions are easily found to be: \begin{itemize}
\item[-] traceless matrices, with eigenvalue $\a = \f{1}{2}\left(\f{1}{\lt_p} - \f{4}{\pi^2}\right)$,
\item[-] matrices proportional to the identity, with eigenvalue $\b = \f{1}{2}\left(\f{1}{\lt + \lt_p} - \f{4}{\pi^2}\right)$. 
\end{itemize}
This suggests that the trivial solution becomes unstable towards traceless perturbations at $\lt_p~\geq~\pi^2/4$ and towards trace perturbation at $\lt+\lt_p ~\geq~ \pi^2/4$. 
Therefore, the following two particular non-zero solutions are examined:\footnote{Other solutions are possible, as in appendix D of \cite{Benedetti:2017fmp}. In that case it was possible to show that such solutions are never global minima of the potential; here the analysis is more complicated and we limit ourselves to conjecture that the analysis of the following two types of solutions suffices to understand the full phase diagram of the model.}
\begin{itemize}
\item $\mu_i = \mu\neq 0 ~\forall i$: The potential takes the form
\begin{equation}
\label{eq:uniform potential}
\f1N V_{\rm eff}(\mu)=\frac{\mu^2}{4\lt_p}\left(1 + \frac{b}{1-b} - \frac{8\lt_p}{3\pi^2}\right) + \frac{2}{3\pi^2}\abs{\mu}^3\arctan{\frac{1}{\abs{\mu}}} - \frac{1}{3\pi^2}\log(1+\mu^2) \;,
\end{equation}
and the equation of motion that $\mu$ must satisfy is
\begin{equation}
\label{eq: eom uniform}
\mu \arctan{\frac{1}{\mu}} = 1 - \frac{\pi^2}{4(\lt + \lt_p)} \;.
\end{equation}
The range of values of the left-hand side tells us that such a solution exists only for $\lt + \lt_p \geq \pi^2/4$.
\item $\Tr \Mt= 0$: Then the potential reduces to a sum over the eigenvalues. We obtain
\begin{equation}
\label{eq:traceless potential}
V_{\rm eff}[\Mt]=\sum_i v(\mu_i)\;,
\end{equation}
where
\be
v(\mu) = \frac{\mu^2}{4\lt_p}\left(1 - \frac{8\lt_p}{3\pi^2}\right) + \frac{2}{3\pi^2}\abs{\mu}^3\arctan{\frac{1}{\abs{\mu}}} - \frac{1}{3\pi^2}\log(1+\mu^2) \;.
\ee
The equation of motion for $\m_i$ is
\begin{equation}
\label{eq: eom traceless}
\mu_i \arctan{\frac{1}{\mu_i}} = 1 - \frac{\pi^2}{4\lt_p} \;.
\end{equation}
The range of the left-hand side tells us that such a solution exists only for $\lt_p \geq \pi^2/4$.
Furthermore, being an even function, monotonic on each semiaxis, there are only two solutions $\mu_i = \pm \t$. The tracelessness condition finally tells us that the two must come in equal number (for odd $N$ we necessarily have either a zero eigenvalue or a violation of the tracelessness condition, which amounts to a subleading effect in $1/N$).
\end{itemize}

Using the equations of motion, we need to compare the values of the potential at the above critical points:
\begin{equation}
V_{\rm eff}(q) = \frac{\tau(q)^2}{12 q} -\frac{1}{3\pi^2}\log(1 + \tau(q)^2),
\end{equation}
with $q = \lt+\lt_p$ in the uniform case and $q=\lt_p$ in the traceless one, and $\tau(q)$ being the solution of $\t \arctan{(1/\t)} = 1 - \pi^2/(4 q)$. Since $\tau(q)^2$ is a monotonically increasing function, and since as a function of $q$, $V_{\rm eff}(q)$ is decreasing monotonically starting from $0$ (the trivial solution), we conclude that
\begin{equation}
\lt_p > \lt+\lt_p \implies V_{\rm eff}(q_{traceless})<V_{\rm eff}(q_{uniform})
\end{equation}
and reciprocally. In other words, the traceless solution wins over the uniform one for $\lt<0$, while the uniform wins for $\lt>0$.

Such transition can be qualitatively understood in terms of the double-trace term: we see that if $\lt<0$, then the double-trace term comes with a positive sign and has to be minimized, showing why the traceless solution wins (when it exists, i.e.\ for $\lt_p>\pi^2/4$), while if $\lt>0$, then the coefficient of the double-trace term is negative and has to be maximized, leading to the uniform solution.

At last, the phase diagram is as shown in Fig.~\ref{fig:phase diagram}.
%%%%%%%%%
%%%%%%%%%
Knowing the value of the potential at the different vacua, we see that the transition from red and blue to green are continuous, hence of second order. Indeed, comparing eq. \eqref{eq: eom uniform} with eq. \eqref{eq: eom traceless}, we see that taking the limit $\lt_p+\lt$ or $\lt_p$ to $\pi^2/4$, $\mu$ and $\tau$ decrease monotonically to zero. 
On the other hand, a first order transition separates the two non-trivial phases: at $\lt = 0$ and $\lt_p>\pi^2/4$, they both have same potential energy (but are distinct), and as the coupling $\lt$ grows or decreases, the uniform or traceless solution become global minima, respectively.

%%%%%%%%%%%%%%%%%%%%%%%%%%%%%
\subsection{Schwinger-Dyson equations in the traceless phase}
\label{sec:SD1bis}
%%%%%%%%%%%%%%%%%%%%%%%%%%%%%

In Sec.~\ref{sec:SD1} we derived the gap equation and beta functions from the Schwinger-Dyson equations in the $U(N)\times U(N^2)$-symmetric phase. Although we expect the beta functions to be independent of such choice it is instructive to do an explicit check, now that we discovered a broken phase.

Assuming that the two-point function breaks the $U(N)\times U(N^2)$ symmetry into $U(N/2)^2\times U(N^2)$, with the following ansatz:
\begin{gather}
M_1 = m (\mathbb{1}_N- 2\mathbb{P}_N), 
\quad \mathbb{P}_N = \begin{pmatrix}
\mathbb{0}_{N/2} & \mathbb{0}_{N/2} \\
\mathbb{0}_{N/2} &\mathbb{1}_{N/2}\end{pmatrix},\\ \label{eq:broken-2pt-ansatz}
G^{-1}(p) = i\slashed{p} \,\mathds{1}_1\otimes \mathds{1}_2 \otimes \mathds{1}_3 + M_1\otimes \mathds{1}_2 \otimes \mathds{1}_3 \;,
\end{gather}
the Schwinger-Dyson equation becomes
\begin{equation}
\label{eq:SDE U(N)-broken}
G^{-1}(p) =  i\slashed{p} \mathds{1}_1\otimes \mathds{1}_2 \otimes \mathds{1}_3 + 2  \f{\l_p}{N^2}\int\f{\dd[3]q}{(2\pi)^3}  (\Tr_{\backslash 1} \tr \left[ G(q)  \right])\otimes \mathds{1}_2 \otimes \mathds{1}_3\;,
\end{equation}
where $\Tr_{\backslash c}$ is a trace on all indices except the one of color $c$.
The coupling $\l$ is missing from the equation because it multiplies a full trace of $G(q)$, which is zero for the ansatz above.
Forgetting momentarily the trace over the $\g$-matrices, we have
\begin{equation}
\Tr_{\backslash 1} \left[G(q)\right] = N^2 \f{-i\slashed{p} + M_1}{p^2+m^2}\;,
\end{equation}
such that the Schwinger-Dyson equation reduces to
\begin{equation}
M_1 = 2\l_p \int \f{\dd[3]q}{(2\pi)^3} \f{\tr(-i\slashed{p} + M_1)}{p^2+m^2}\;,
\end{equation}
or the mass gap equation
\begin{equation}
m = 8\l_p\int \f{\dd[3]q}{(2\pi)^3} \f{m}{p^2+m^2}\;,
\end{equation}
that is, nothing more than eq.~\eqref{eq:gap-1c} with $\l = 0$, thus leading directly to \eqref{eq:beta-1c}. 

%%%%%%%%%%%%%%%%%%%%%%%%%%%%%
\subsection{Anomalous dimension}
%%%%%%%%%%%%%%%%%%%%%%%%%%%%%

We have three non-trivial fixed points (out of which one has two relevant directions while the others have one), and for each of them we can compute the conformal dimension of the intermediate field. It happens, that in all cases, the computation is almost unchanged and gives the same result:

\begin{itemize}

\item $(\lt,\lt_p)=(\pi^2/4,0)$

This is the UV fixed point of the usual GN model. The limit $\l_p\to 0$ constrains to zero the traceless part of the intermediate field \cite{Benedetti:2017fmp}, and thus
it is equivalent to starting from the action \eqref{eq:GNY}. Integrating out the fermions:
\be
S_{\rm int}[\s] = N^3 \int d^3 x \;\left( \frac{1}{4\l}\sigma^2 - \log\left(\slashed{\partial} + \s \right)\right).
\ee
At the fixed point, $\la \s\ra =0$ and the inverse propagator is obtained by the second functional derivative with respect to $\s$, computed at $\s=0$, i.e.:
\begin{equation} \label{eq:effective_prop_case1}
\f{1}{N^3}\la\s(p)\s(-p)\ra^{-1} = \frac{1}{2\l} - \tr \int\frac{\dd[3]q}{(2\pi)^3}\frac{\slashed{q}(\slashed{q}-\slashed{p})}{q^2(q-p)^2}
=\frac{1}{2\l} - 4 \int\frac{\dd[3]q}{(2\pi)^3}\frac{q^2-q\cdot p}{q^2(q-p)^2}.
\end{equation} 

The last integral can be computed as
\be
\begin{split}
4 \int\frac{\dd[3]q}{(2\pi)^3}\frac{q^2-q\cdot p}{q^2(q-p)^2} &= \f{2}{\pi^2} \int_0^\L dq - \f{2p^2}{(2\pi)^2} \int_0^{+\infty} dq \int_{-1}^{+1} d(\cos\theta) \f{1}{q^2+p^2-2 q p \cos\theta}\\
&= \f{2}{\pi^2} \L - \f{p}{4} \;.
\end{split}
\ee
The linear divergence is cancelled by the fixed point condition $\l = \frac{\pi^2}{4\Lambda}$, thus yielding
\be
\la\s(p)\s(-p)\ra =  \f{4}{N^3 p} \;,
\ee
corresponding to a conformal dimension $\D_\s=1$, which is also the dimension of $\psib_{abc}\psi_{abc}$.

\item $(\lt,\lt_p)=(0,\pi^2/4)$

Let us recall the effective action of the intermediate field 
\begin{align}
S_{\rm int}[M] = &\int \frac{1}{2}M^*_{ij}K_{ij;kl}M_{kl} - N^2 \tr \Tr\left[\log\left(\slashed{\partial} + \frac{\sqrt{2\l_p}}{N}M\right)\right],\\
&K_{ij;kl} = \delta_{ik}\delta_{jl} + \frac{b}{(1-b)N}\delta_{ij}\delta_{kl}.
\end{align}
It is convenient to introduce again the rescaled matrix $\Mt = \frac{\sqrt{2\l_p}}{N}M$. Derivating twice with respect to an eigenvalue $m_i$ of $\Mt$, and setting $b=0$, gives, after a Fourier transform
\begin{equation}
\f{1}{N^2} \frac{\delta^2 S_{\rm int}}{\delta m^2}\big|_{m=0} = \frac{1}{2\l_p} -\tr \int\frac{\dd[3]q}{(2\pi)^3}\frac{\slashed{q}(\slashed{q}-\slashed{p})}{q^2(q-p)^2}\;,
\end{equation} 
namely the same expression as \eqref{eq:effective_prop_case1}, with $\l_p$ replacing $\l$.
The linear divergence is cancelled by the fixed point condition, $\frac{1}{2\l_p} = \frac{2\Lambda}{\pi^2}$, and we arrive at the same propagator (hence same conformal dimension) as in the usual three-dimensional Gross-Neveu model.

\item $(\lt,\lt_p)=(-\pi^2/4,\pi^2/4)$

Here, because $b=1$ is a singular point of $K$, we need to take a few steps back \cite{Benedetti:2017fmp}. The fermionic interaction action was written with a matrix-like field $B_{ij} = \bar{\psi}_{ibc}\psi_{jbc}$ as
\begin{align}
&S_{\rm int} = -\frac{\l_p}{N^2}\int B^*_{ij}C_{ij;kl}B_{kl}\;,\\
C_{ij;kl} = \delta_{ik}\delta_{jl} - \frac{b}{N}\delta_{ij}\delta_{kl} &= (\bm{1} - \bm{P})_{ij;kl} + (1 - b)\bm{P}_{ij;kl}\;, \qquad \bm{P}_{ij;kl} = \frac{\delta_{ij}\delta_{kl}}{N}\;.
\end{align}
Since $\bm{P}$ projects on the trace part of the matrix, it appears clearly that $b=1$ restricts us to work with a traceless $B$. Except for this constraint on the fields (which then follows for the matrix-like intermediate field $M$ \footnote{The trace of $M$ couples to that of $B$ and its effective action will be identical to eq. \eqref{eq:S_M}, except for an absent double-trace term.}), the computation of the effective propagator will be identical to other two cases above.
\end{itemize}

%%%%%%%%%%%%%%%%%%%%%%%%%%%%%
\section{$U(N)\times U(N)\times U(N)$-symmetric model}
\label{S:2}
%%%%%%%%%%%%%%%%%%%%%%%%%%%%%

Adding to the model defined in \eqref{eq:action}-\eqref{eq:S_int1} other pillow interactions which differ by simultaneous permutations of the three tensor indices of all the fields, we break the $U(N)\times U(N^2)$ symmetry down to $U(N)\times U(N)\times U(N)$. In the tensor model literature it is usual to refer to an index location (first, second, or third index, in our case) as a color, and hence such permutations of indices are called color permutations. There are three distinguishable colorings for the pillow interaction, one for each choice of {\emph{transmitted color}}, i.e.\ for each choice of index being associated to the vertical lines of Fig.~\ref{fig:interactions}. Considering that of course there is only one coloring for the double-trace interaction, we have in general four independent couplings. We will restrict the theory space by demanding \emph{color symmetry} of the action, i.e. invariance under permutations of the indices, thus writing for the new interacting part of the action
\be \label{eq:S_int123}
S_{\rm int}[\psi,\psib]= -\frac{\l}{N^3}  \int d^3 x \; (\bar{\psi}_{abc}\psi_{abc})^2 - \frac{\l_p}{N^2} \sum_{\ell =1}^3  \mathcal{P}_\ell[\psi,\psib] \;,
\ee
where $\mathcal{P}_\ell[\psi,\psib]$ is the pillow interaction with transmitted color $\ell$.

%%%%%%%%%%%%%%%%%%%%%%%%%%%%%
\subsection{Schwinger-Dyson equations and $\beta$-functions}
\label{sec:SD123}
%%%%%%%%%%%%%%%%%%%%%%%%%%%%%

\begin{figure}
\centering 
\includegraphics[width=0.7\textwidth]{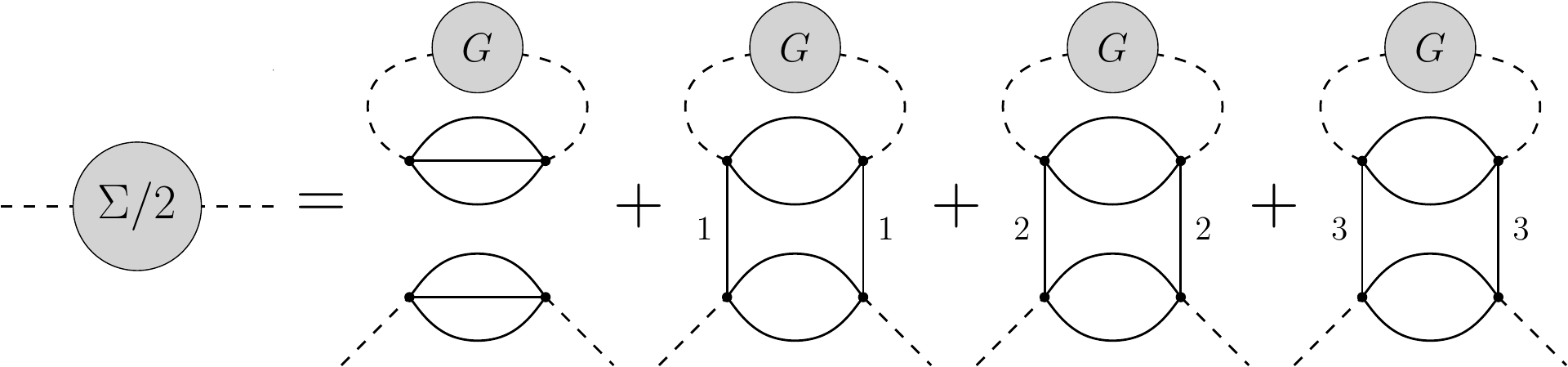}
\caption{\label{fig:SDE-3c}Schwinger-Dyson equation at large $N$ for the self-energy.}
\end{figure}

\begin{figure}
\begin{minipage}{0.5\textwidth}
  	\centering 
	\includegraphics[width=0.8\textwidth]{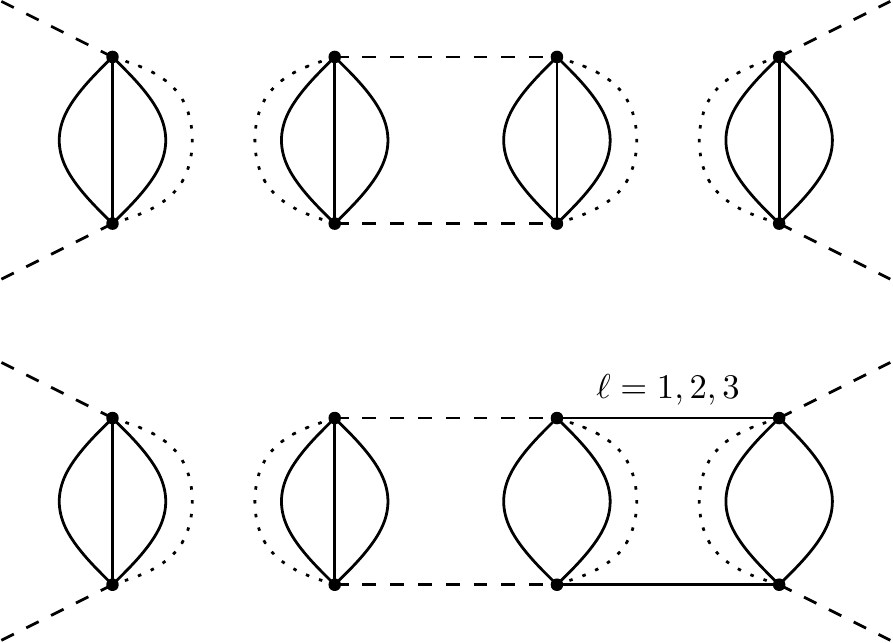}
        \end{minipage}
\begin{minipage}{0.5\textwidth}
      	\centering
	\includegraphics[width=0.8\textwidth]{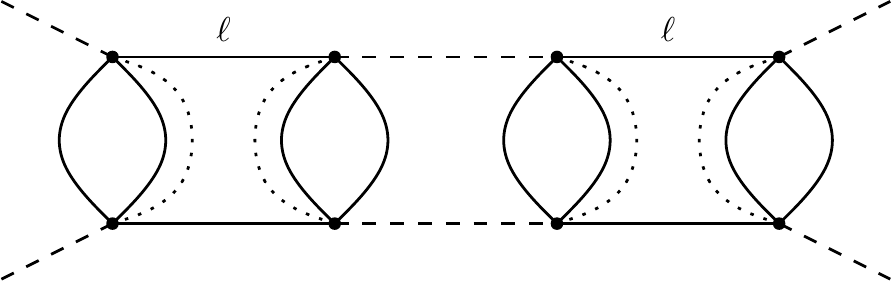}\\	
	\vspace{.6cm}
	\includegraphics[width=0.8\textwidth]{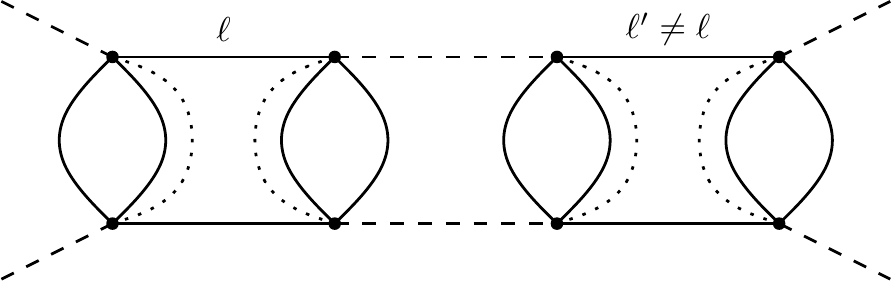}
        \end{minipage}
\caption{\label{fig:4-point graphs-3c}Leading order graphs renormalizing the couplings $\lt_p$ (top right) and $\lt$ (all the others).}
\end{figure}

Following \cite{Benedetti:2017fmp}, the SD equations in momentum space write as \footnote{$\Tr$ is a trace on all the color indices, while with $\Tr_{\backslash c}$ the color $c$ is not traced on. As before, $\tr$ is a trace on the $\g$-matrix space.}
\begin{equation}
\label{eq:SDE-3c}
G^{-1}(p) = G^{-1}_0(p) + 2 \int\f{\dd[3]q}{(2\pi)^3}  \tr \left[\f{\l}{N^3} \Tr G (q) + \f{\l_p}{N^2} \left(\Tr_{\backslash 1} G(q)  + \Tr_{\backslash 2} G(q) +\Tr_{\backslash 3} G(q)\right)\right]\;,
\end{equation}
as also depicted in Fig.~\ref{fig:SDE-3c}.
%%%%%%%%%%%%%%%%
%%%%%%%%%%%%%%%%
Assuming that the $U(N)$ symmetry is unbroken by the vacuum, an ansatz for the full propagator is again given by \eqref{eq:2pt-sym-ansatz}, with the diagonal $\hat{G}(p)^{-1}= i\slashed{p} + m $; thus, similarly to subsection \ref{sec:SD1}, the gap equation is
\begin{equation}
m = 2(\l + 3\l_p)\int\frac{\dd[3]q}{(2\pi)^3} \frac{\tr(-i\slashed{q} + m)}{q^2 + m^2}, 
\end{equation}
or
\begin{equation}
\frac{1}{\l+3\l_p}=8 \int_{\abs{p}<\Lambda}\frac{\dd[3]p}{(2\pi)^3}\frac{1}{p^2+m^2} = \frac{4}{\pi^2}\left(\Lambda - m \arctan\left(\frac{\Lambda}{m}\right)\right).
\end{equation}
In terms of the dimensionless couplings ($\lt_p \equiv \Lambda\l_p$ and $\lt \equiv \Lambda\l$) and with $\kappa \equiv 4/\pi^2$, the $\beta$-functions read 
\be
\label{eq:beta-unbroken}
\beta+3\beta_p = \left(\lt + 3 \lt_p\right) - \kappa\left(\lt + 3 \lt_p\right)^2 \;.
\ee
By direct inspection of the one loop diagrams at leading order in $1/N$, depicted in Fig.~\ref{fig:4-point graphs-3c}, we can disentangle the two beta functions, obtaining:
\begin{align} \label{eq:beta-3c}
\beta &=  \lt - \k\left(\lt^2 + 6\lt \lt_p + 6\lt_p^2\right)\;,\\
\beta_p &= \lt_p - \kappa\lt_p^2 \;.
\end{align}

%%%%%%%%%
%%%%%%%%%

With the experience of the previous section, we can also consider the case of broken $U(N)$ symmetry, and disentangle the two beta functions by combination of the symmetric and broken results. 
Anticipating the results of section \ref{sec:potential-3c}, it turns out that there is a broken phase where we can make precisely the same ansatz as in \eqref{eq:broken-2pt-ansatz}. The calculation then proceeds exactly as in that section, after having noticed that in \eqref{eq:SDE-3c} the $ \Tr_{\backslash 2}$ and $ \Tr_{\backslash 3}$ terms vanish because the trace on color 1 is zero.
Therefore, the beta function for $\l_p$ is unchanged, and combining it with that of the unbroken case (eq. \eqref{eq:beta-unbroken}), we find again \eqref{eq:beta-3c}, as expected.

It is also interesting to point out that $\b + 2\b_p = 0$ along $\l+2\l_p=0$.

\begin{figure}
\centering
\includegraphics[width=0.6\textwidth]{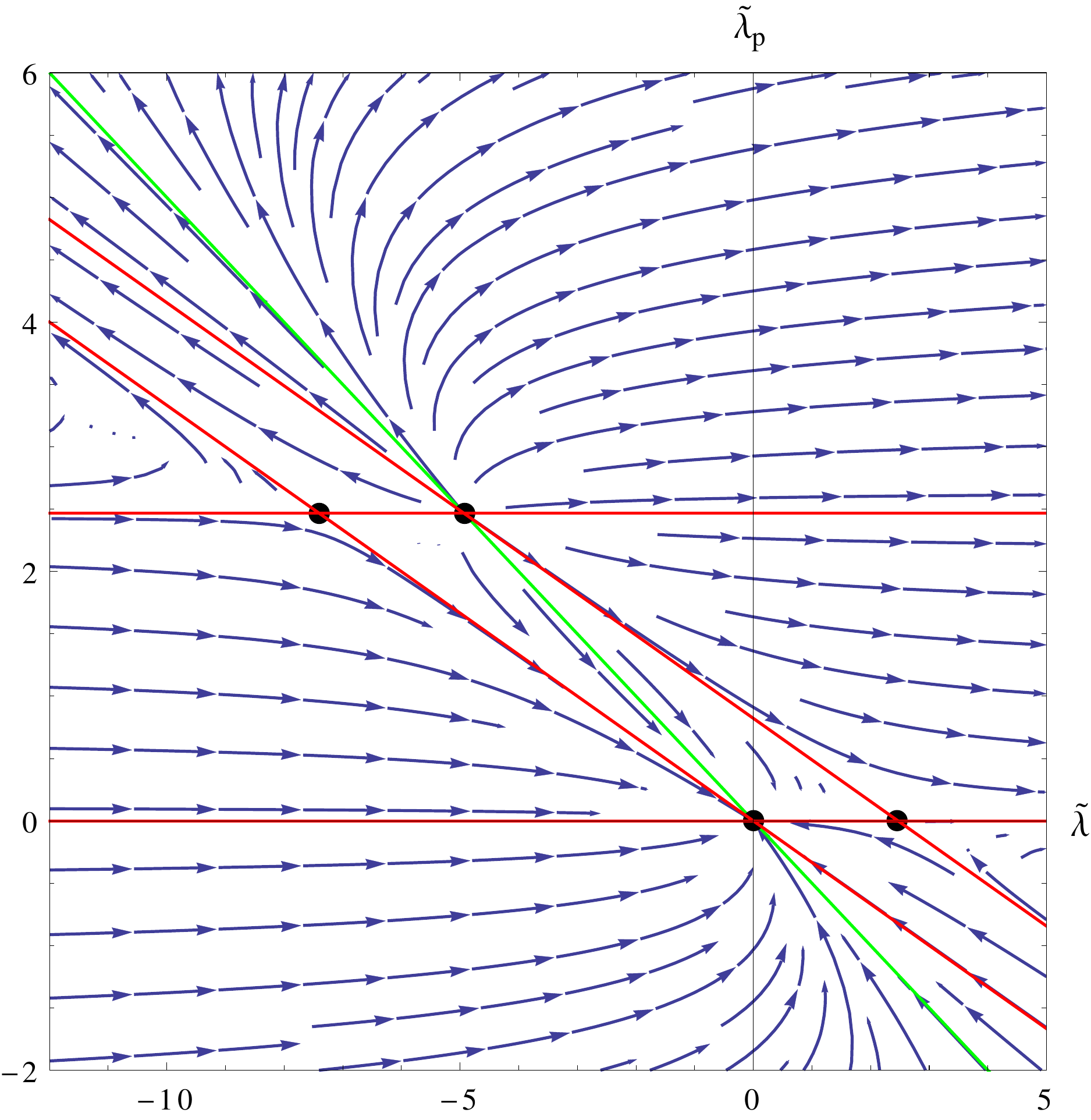}
\caption{Renormalization group flow in the $\left(\lt,\lt_p\right)$-plane. Arrows point towards the $IR$, black dots denote the fixed points, and red lines mark the zeros of either $\beta_p$ or $\beta+3\beta_p$. The green line corresponds instead to $\beta+2\beta_p=0$, which morally replaces the vertical line $\beta=0$ of Fig.~\ref{fig:RG-flow-fermions}.}
\label{fig:RG-flow-3c}
\end{figure}

We picture the resulting flow on Fig.~\ref{fig:RG-flow-3c}, seemingly a distorted version of Fig.~\ref{fig:RG-flow-fermions}.
In fact the critical exponents are also the same as in the previous model.
%%%%%%%%%
%%%%%%%%%

%%%%%%%%%%%%%%%%%%
\subsection{Effective potential and phase diagram}
\label{sec:potential-3c}
%%%%%%%%%%%%%%%%%%

Rewriting all the quartic interactions in terms of intermediate fields, the action takes the following form
\begin{gather}
\label{eq:action_cIF}
S[M_1,M_2,M_3] = \int \frac{1}{2} \sum_{c=1,2,3}\left[\Tr(M_c^2) + \frac{b}{(1-b)N}(\Tr M_c)^2\right] - \tr \Tr\left[ \ln\left(\slashed{\partial} + \frac{\sqrt{2\l_p}}{N} R\right)\right]\\
R \equiv M_1\otimes \mathds{1}_2 \otimes \mathds{1}_3 + \mathds{1}_1 \otimes M_2 \otimes \mathds{1}_3 + \mathds{1}_1 \otimes \mathds{1}_2 \otimes M_3\;; \quad b = -\f{\l}{3\l_p},
\end{gather}
with the three intermediate fields $M_1, M_2, M_3$ needed for the three pillow interaction terms. The original GN-interaction is split into three identical parts, thus in effect changing $\l \rightarrow \l/3$ in each of the $(\Tr M_c)^2$ terms.
Again, coming to adimensional variables, rescaling such that $\Mt_i \equiv N \L^{3/2}M_i, ~ \forall i$, and using the $U(N)$ symmetry to diagonalize, we arrive at
\be
\begin{split}
V &\left[\{\mu_{1,i}, \mu_{2,j} , \mu_{3,k}\}\right] \equiv \f{S\left[\Mt_1,\Mt_2,\Mt_3\right]_{|_{\Mt_i={\rm const.}}}}{N^3\L^3 {\rm Vol}} \\
&\quad = \f{1}{N}\sum_{c} \frac{1}{4\lt_p} \left[\sum_i\mu_{c,i}^2 - \frac{\lt}{(\lt+3\lt_p) N}\left(\sum_i\mu_{c,i}\right)^2 \right]\\
& \qquad \qquad + \f{1}{N^3}\sum_{1\leq i,j,k\leq N}\frac{1}{3\pi^2}\kappa(\mu_{1,i},\mu_{2,j},\mu_{3,k})\;,
\end{split}
\ee
\begin{gather}
\kappa(\mu_{1,i},\mu_{2,j},\mu_{3,k}) = 2\mu^3\arctan\frac{1}{\mu} - 2\mu^2 - \log(1+\mu^2), \\
\mu \equiv \mu_{1,i}+\mu_{2,j}+\mu_{3,k} \;.
\end{gather}
We are looking for the vacua of this potential. The equations of motion for $\mu_{c,i}$ read 
\begin{equation}
\f{1}{2\lt_p}\left[\m_{c,i} - \f{\lt}{(\lt +3 \lt_p)N}\sum_j\m_{c,j}\right] + \f{2}{\pi^2N^2}\sum_{1\leq j,k\leq N}\left[\mu^2\arctan\f{1}{\mu} - \mu\right] = 0.
\label{eq:eom_3c}
\end{equation}

To begin, a useful remark is that the three different color-intermediate fields must have the same trace at the saddle points. Indeed, this is seen by summing eq. \eqref{eq:eom_3c} over $i$. Because the second term in squared brackets depends only on $\mu$, all colors end up with the same equation (of the type $\Tr[\Mt_c]=F[\{\m\}]$, with the same right-hand side), hence the equality of traces. 

Another straightforward observation is that $M_i = 0 ~ \forall i$ is always a solution. In order to find other solutions, it is helpful to search for unstable directions around this point, i.e. analyse the Hessian of the potential. 

Developing around the point $R = 0$ allows to discard the $\arctan$ term, of higher order. The first derivative gives\footnote{The exponent $(c)$ in $\d_{ij}^{(c)}$ only serves to keep track of what color the indices belong to.}
\begin{gather}
\pdv{V}{M_{c,ij}} = (A +C) M_{c,ji} + B \f{\Tr M_{c}}{N} \d_{ij}^{(c)} + \sum_{c^\prime \neq c} C \f{\Tr M_{c^\prime}}{N} \d_{ij}^{(c)}\; ,\\
A = \f{1}{2\lt_p}\; ,\quad B = \f{b}{2\lt_p(1-b)}\; , \quad C = - \f{2}{\pi^2}\;.
\end{gather}
The second derivative is 
\begin{equation}
 \pdv{V}{M_{c,ij}}{M_{c^\prime,kl}} = \d_{cc^\prime}\left((A + C) \d_{ik}\d_{jl} + B\f{\d_{ij}\d_{kl}}{N}\right) + C\f{\d_{ij}^{(c)}\d_{kl}^{(c^\prime)}}{N},
\end{equation}
and can be rewritten as a matrix in color-space
\begin{gather}
H = 
\begingroup
\renewcommand*{\arraystretch}{1.5}
\begin{pmatrix}
\a(1 - P_1)+ \b P_1 & C \f{\Pi_1\Pi_2}{N} &  C \f{\Pi_1\Pi_3}{N}\\
C \f{\Pi_2\Pi_1}{N} &\a(1 - P_2)+ \b P_2 & C \f{\Pi_2\Pi_3}{N}\\
C \f{\Pi_3\Pi_1}{N} & C \f{\Pi_3\Pi_2}{N} &\a(1 - P_3)+ \b P_3
\end{pmatrix},
\endgroup \\
\a = A + C\; \quad \b = A + B + C.
\end{gather}
We defined the projector $P_c$ on the trace of a matrix of given color $c$
\begin{equation}
P_{c;ij,kl} \equiv \frac{1}{N}\Pi_{c,ij}\Pi_{c,kl}, \quad \Pi_{c,ij}\equiv \d_{ij}^{(c)}.
\end{equation}
Articulated as such, the Hessian $H$ is easy to diagonalize\footnote{Using a simplified notation in which, for instance, an element $a$ on the second line of the column vector is intended to represent the element $\mathbb{1} \otimes a \otimes \mathbb{1}$, and analogously with the other lines.}:
\begin{itemize}
\item we have the set of eigenvectors associated to traceless matrices, of form 
\begin{equation}
E^1 = \begin{pmatrix}Q\\0\\0\end{pmatrix}, \quad E^2 = \begin{pmatrix}0\\Q\\0\end{pmatrix},\quad E^3 = \begin{pmatrix}0\\0\\Q\end{pmatrix}, \quad \Tr Q = 0,
\end{equation}
with eigenvalue $\alpha = \f{1}{2}\left(\f{1}{\lt_p} - \f{4}{\pi^2}\right)$ ; 
\item the eigenvector associated to matrices proportional to the identity, of the form
\begin{equation}
E^s = \g\begin{pmatrix}1\\1\\1\end{pmatrix},
\end{equation}
with eigenvalue $\b + 2C = \f{3}{2}\left(\f{1}{\lt + 3\lt_p} - \f{4}{\pi^2}\right)$ ; 
\item and finally the are eigenvectors
\begin{equation}
e^1 = \g_1 \begin{pmatrix}1\\-1\\0\end{pmatrix}, \quad e^2 = \g_2 \begin{pmatrix}1\\0\\-1\end{pmatrix},
\end{equation}
with eigenvalue $\b - C = \f{3\lt_p}{\lt + 3\lt_p}$.
\end{itemize}
The first two correspond to instabilities that decrease the potential (in the ranges where the eigenvalues become negative), while the last always increases it for the range of couplings allowed by the requirement of boundedness of the potential (and in addition they do not satisfy the equations of motion, because the traces of the three matrices are not equal).

\begin{figure}
\centering
\includegraphics[width=0.7\textwidth,height=0.48\textwidth]{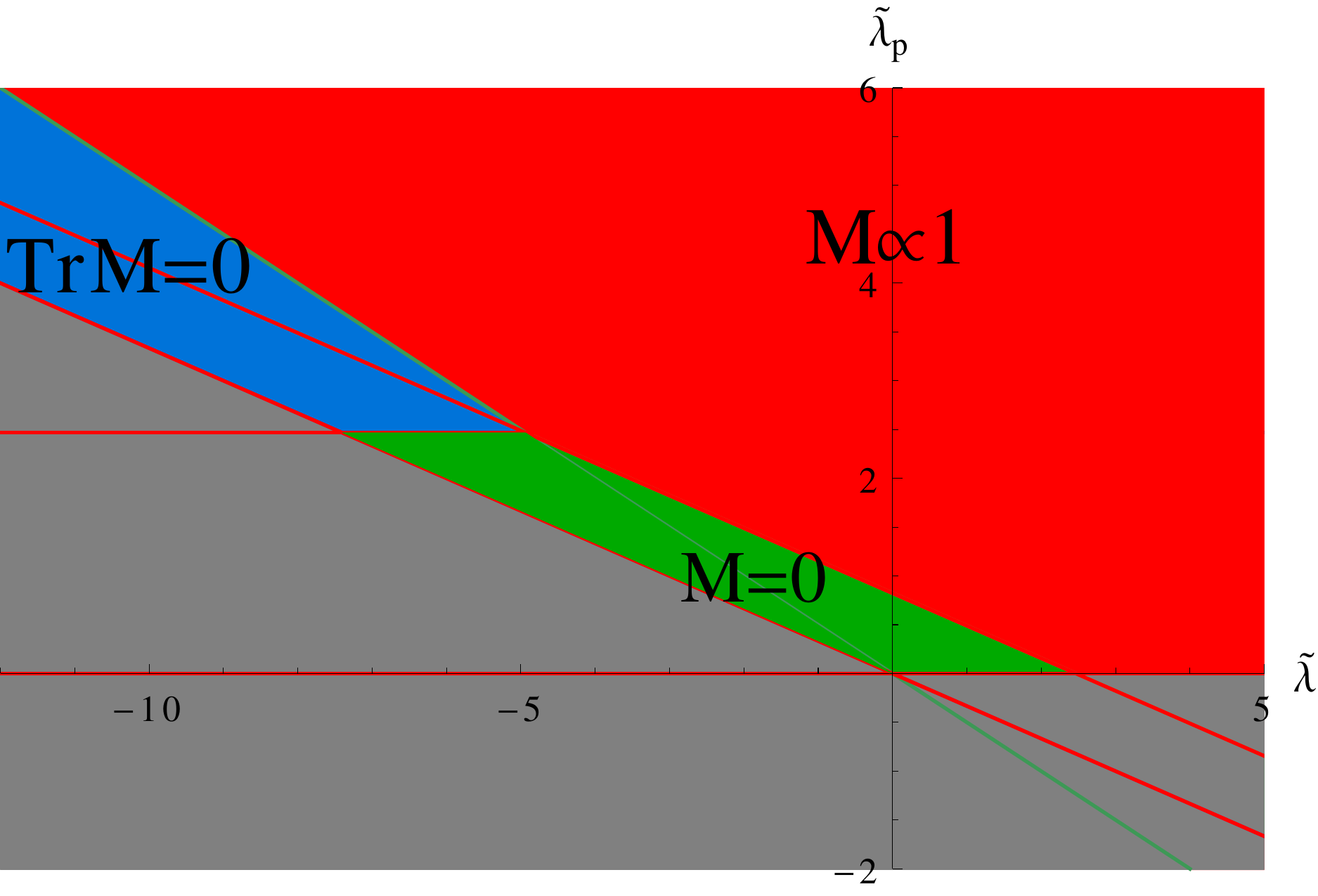}
\caption{Phase diagram of the color-symmetric fermionic TGN model in 3$d$. The color code is the same as in Fig. \ref{fig:phase diagram}.}
\label{fig:phase-diagram-3c}
\end{figure}

In light of this analysis, we can constrain the form of the intermediate field in our search of new minima of the potential: 
\begin{itemize}
%%%%
\item Assuming $M_i = m\mathbb{1}_N ~ (i=1,2,3)$, the equations of motion imply for $m$ the equation
\be
1 - 3m\arctan\f{1}{3m} = \f{\pi^2}{4} \f{1}{\lt+3\lt_p},
\label{eq:constraint-m}
\ee
allowing such solutions to exist only for $\lt+3\lt_p>\pi^2/4$. After using eq. \eqref{eq:constraint-m} to remove the $\arctan$ term, their potential energy is given by
\be
V[m] = \f{1}{12 (\lt+3\lt_p)}(3m)^2 - \f{1}{3\pi^2}\log\left(1+(3m)^2\right),
\label{eq:potential-m-3c}
\ee
corresponding to a non-trivial minimum as soon as $\lt + 3\lt_p>\pi^2/4$.
%%%%
\item Assuming that the intermediate fields are traceless reduces the equations of motion \eqref{eq:eom_3c} to the following:
\begin{equation}
\mu_{1,i} \left(\f{\pi^2}{4\lt_p} - 1\right) + \f{1}{N^2}\sum_{jk}(\mu_{1,i}+\mu_{2,j}+\mu_{3,k})^2\arctan\f{1}{\mu_{1,i}+\mu_{2,j}+\mu_{3,k}}=0\; .
\label{eq:eom_traceless}
\end{equation}
Taking only one non-trivial matrix $M_1 = w(\mathbb{1}_N-2 \mathbb{P}_N)$, and $M_2 = M_3 = \mathbb{0}_N$, $w$ is hold by eq. \eqref{eq:eom_traceless} to 
\be 
1 - w \arctan\f{1}{w} = \f{\pi^2}{4\lt_p}\;,
\label{eq: single-traceless-eom}
\ee
while the potential energy is 
\be
V[w] = \f{1}{12 \lt_p}w^2 - \f{1}{3\pi^2}\log\left(1+w^2\right),
\label{eq:potential-w-3c}
\ee
again corresponding to a non-trivial minimum for $\lt_p>\pi^2/4$.

The cases of two and three traceless matrices are examined in App. \ref{app:traceless_matrices}, where we show in detail that the single-traceless potential dominates double- and triple-traceless solutions above $\lt_p = \pi^2/4$.
\end{itemize}
The potentials of eq. \eqref{eq:potential-m-3c} and \eqref{eq:potential-w-3c} compare easily: $m$ and $w$ are controlled by similar equations, and they grow monotonically with the coupling. Consequently, we have that $V[w]<V[m]$ for $\lt + 3\lt_p>\lt_p$ or $\lt+2\lt_p>0$. In other words, the green line outlined in Fig. \ref{fig:RG-flow-3c} and \ref{fig:phase-diagram-3c} sets apart the traceless vacuum from the traceful one.

\

Let us summarize the results.
%%%%%%%%%
%%%%%%%%%
Integrating out the fermions, we obtained the effective potential of the intermediate fields. The associated equations of motion were too difficult to solve in full generality. Nevertheless, after a stability analysis around the trivial vacuum, showing that the unstable perturbations beyond some critical lines are given by traceless and pure trace modes, we explored those two types of solutions (i.e.\ traceless matrices and matrices proportional to the identity) and showed in what range of the couplings they minimized the energy. The conclusion is very similar to that of the previous section. Indeed, three phases appear: the trivial, traceful or traceless solutions dominate in turn, as displayed on the phase diagram of Fig. \ref{fig:phase-diagram-3c}. However, in the last case, a startling new feature is the breaking of color-symmetry. The transitions between trivial and traceful or traceless are continuous, while a discontinuous transition separates the non-trivial phases.

%%%%%%%%%%%%%%%%%%%%%%%%%%%%%
\section{Conclusions}
%%%%%%%%%%%%%%%%%%%%%%%%%%%%%
In this paper we have studied the large-$N$ limit of two fermionic quantum field theories with symmetry group $U(N)\times U(N^2)$ and $U(N)\times U(N)\times U(N)$, respectively. The study of such type of models is motivated on one hand by the attempt of generalizing the behaviour found in the SYK model to higher dimensional models and on the other by the possibility of generalizing the Klebanov-Polyakov duality between vector models at large $N$ and higher-spin theories.
More generally, such models are also of interest as they provide a new solvable limit in quantum field theory.

The specific three-dimensional quartic tensor models we focused on have the simplifying feature of admitting a rewriting in terms of matrix intermediate fields. We studied their RG flow, and in addition to the trivial IR fixed point and to the standard UV fixed point of the Gross-Neveu model (where the tensorial interactions that break the $U(N^3)$ symmetry are zero), we found two new fixed points: one with two relevant directions, and another with one relevant and one irrelevant direction. We identified three phases of the vacuum by analyzing the effective potential of the intermediate field: a massless phase preserving all the symmetries of the model; a phase with dynamically generated mass, breaking only the chiral symmetry; and a phase breaking the chiral symmetry, as well as one of the $U(N)$ subgroups of the symmetry group (and as a consequence breaking also the color symmetry, when present in the original model). We found that the conformal dimension of the unconstrained component of the intermediate field (and thus of the corresponding composite operator $\psib \psi$) is the same at all the nontrivial fixed points, $\D_{\psib \psi}=1$ (to be compared to the Gaussian fixed point value, $\D_{\psib \psi}=2$), thus matching that of the critical vector GN model.
The dimensions of the (integrated) quartic operators are instead different, as they change from relevant to irrelevant from one fixed point to another, but are the same in modulus, which is always equal to one.

As discussed in \cite{Benedetti:2017fmp}, the Goldstone bosons of the phase with broken $U(N)$ subgroup are governed by a complex Grassmannian non-linear sigma model, more precisely with Grassmannian $\mathrm{Gr}(N,N/2)\equiv U(N)/(U(N/2)\times U(N/2))$. Although we have not made further use of this fact here, it is interesting to notice the appearance of such models in the context of tensor models, something that might deserve further study.\footnote{A different pattern of spontaneous symmetry breaking has recently been discussed in \cite{Diaz:2018eik}, in which a $U(N)^2$ subgroup of the symmetry group gets broken down to its diagonal subgroup $U(N)$.}

Among the many possible extensions of the model, two are particularly worth mentioning.

In order to better understand the differences between tensor and vector or matrix models, a gauged version of the above is under investigation. 
As in the vector case, one motivation for gauging these models is to restrict the state space to singlets, the restriction required for a holographic mapping to higher spin theory makes sense \cite{Aharony:2011jz,Giombi:2011kc}.
Therefore, we hope that gauging our tensorial models could serve as a concrete starting point to connect with the duality recently proposed by Vasiliev \cite{Vasiliev:2018zer}. It would also be interesting to study the persistence of the phase with broken $U(N)$ symmetry, and the associated Higgs mechanism.

Lastly, it would be interesting to further study models with different symmetry groups, allowing the complete-graph (or tetrahedron) quartic interaction. Finding healthy nontrivial fixed points for fermionic systems in the presence of such interaction has so far remained challenging (see \cite{Benedetti:2017fmp,Prakash:2017hwq}), but it is worth persevering, as such models would have higher chance of displaying genuinely new physical behavior.

\section*{Acknowledgements}

\noindent We would like to thank Razvan Gurau, Timoth\'e Poulain, and Vincent Rivasseau for useful discussions.

\newpage

%%%%%%%%%%%%%%%%%%%%%%%%%%%%%
\appendix
\section{\texorpdfstring{$\gamma$}{g}-matrices in odd-dimensions}
\label{sec:appendix gamma}
%%%%%%%%%%%%%%%%%%%%%%%%%%%%%
Construction irreducible representations of the Clifford algebra in any dimension $D$
\begin{equation}
\{\g^\mu,\g^\nu\} = 2\eta^{\mu\nu} \;,
\end{equation}
is quite standard (see for instance \cite{Park:2005cyw}). Even dimensions admit a unique irreducible representation, with matrices of dimension $2^{D/2}$, that can be constructed with the following cyclic structure\footnote{All definitions below are correct up to $i$ factors required accord ing to the signature.}:
\begin{align}
\g^1 &\equiv \s_1\otimes \mathbb{1}\otimes \mathbb{1}\dots\\
\g^2 &\equiv \s_2\otimes \mathbb{1}\otimes \mathbb{1}\dots\\
\g^3 &\equiv \s_3\otimes \s_1 \otimes \mathbb{1}\dots\\
\g^4 &\equiv \s_3\otimes \s_2 \otimes \mathbb{1}\dots\\
\g^5 &\equiv \s_3\otimes \s_3 \otimes \s_1\dots, 
\end{align}
and so on, stopping at $D$. 
There, one can also introduce an extra hermitian matrix, squaring to one and anticommuting with $\g^i ~ (1\leq i \leq D)$
\begin{equation}
\g^{(D+1)} \equiv \g^1\cdots\g^D\;.
\end{equation} 
Irreducible representations in odd dimensions $D$ use the same $2^{\f{D-1}{2}}$ dimensional representation as in $D-1$ dimensions taking $\g^{D} = \pm\g^{(D)}$ as the last needed to complete the set. Two inequivalent representations exist, differing by this ``$\pm$'' sign.

In the main text, it was convenient to work with a reducible representation of the Clifford algebra, in order to define a chiral transformation in terms of $\g_5$. We use the following realization of the three dimensional Euclidean Clifford algebra. The $\gamma$-matrices write as
\begin{equation}
\gamma^\mu  = 
\begin{pmatrix}
\tilde{\gamma}^\mu & 0 \\
0 & -\tilde{\gamma}^\mu\end{pmatrix} \quad 
\tilde{\gamma}^1 = \sigma_1 = 
\begin{pmatrix}
0 & 1\\
1& 0 \end{pmatrix} \quad
\tilde{\gamma}^2 = \sigma_2 = 
\begin{pmatrix}
0 & -i\\
i & 0\end{pmatrix} \quad
\tilde{\gamma}^3 = \sigma_3 = 
\begin{pmatrix}
1 & 0\\
0 & -1\end{pmatrix}\;.
\end{equation}
The fermion fields write as
\begin{equation}
\psi = \begin{pmatrix}\psi_1\\\psi_2\end{pmatrix},
\end{equation}
where $\psi_i$'s are 2-component Dirac spinors. Finally our choice for a ``$\gamma^5$''-matrix, such that $\{\gamma^\mu,\gamma^5\}=0$ and $(\g^5)^2 = \mathbb{1}$, is
\begin{equation}
\gamma^5 = \begin{pmatrix}
& \mathbb{1}\\
\mathbb{1} & \end{pmatrix}.
\end{equation}

%%%%%%%%%%%%%%%%%%%%%%%%%%%%%
\section{Other representations of the $U(N)\times U(N^2)$-symmetric model}
\label{app:rectangular}
%%%%%%%%%%%%%%%%%%%%%%%%%%%%%

Given the rectangular matrix structure of the $U(N)\times U(N^2)$ model of Sec.~\ref{sec:S2}, one might wonder why we perform the intermediate field analysis on the index taking values from one to $N$ and not on the one taking values from one to $N^2$, and whether choosing the second option we would discover also a spontaneous symmetry breaking of the $U(N^2)$ subgroup.
The intuitive reason why such option is not favorable is that we would have an intermediate field which is an $N^2\times N^2$ matrix, thus containing more variables than the original fermionic field. We would then expect such formulation to contain redundant information.

In order to illustrate this in a simplified context, let us consider the zero dimensional bosonic analogue of our model, with only one pillow interaction and no double-trace, i.e.\ the model studied in \cite{Benedetti:2015ara}, which however we now rewrite in rectangular matrix notation.\footnote{For tensor models with a rectangular-matrix interpretation see also \cite{doubletens,Bonzom:2013lda}.} The model is defined by the partition function
\be \label{eq:Z-0d}
Z(g) = \int [d\vph d\vphb] e^{- N^2\left( \vphb_{aA}\vph_{aA}-\f{g}{2} \vphb_{aA}\vph_{a'A}\vphb_{a'A'}\vph_{aA'}\right)} \;,
\ee
where the measure is normalized with respect to the free theory, as usual we have summation on the repeated indices, and with respect to \eqref{eq:action} we have rescaled the fields by $N$ to pull out an $N^2$ in front of the action.
Notice that the action can be viewed as a square-matrix action for either
\be
\phi_{ab} \equiv \vph_{aA}\vphb_{bA}
\ee
 or 
\be
\Phi_{AB}\equiv \vphb_{aA}\vph_{aB}\;.
\ee 
Naively, one would expect an action of order $N^3$ in the first case, and an action of order $N^4$ in the second, as the action writes in terms of single traces of square matrices of size $N$ and $N^2$, respectively. However, being one and the same model, the free energy must be the same in the two cases. The reason why the naive reasoning fails is that the two square matrices $\phi_{ab}$ and $\Phi_{AB}$ have the same number of non-zero eigenvalues, which are the squares of the non-zero singular-values of the rectangular matrix $\vph_{aA}$.
The singular-value decomposition was applied to rectangular matrix models in \cite{Anderson:1990nw,DiFrancesco:2002mvz}, and we concisely repeat it here. We write
\be
\vph_{aA} = P_{ab} X_{bB} Q^\dagger_{BA} \;,
\ee
where $P\in U(N)$ is a unitary transformation that diagonalizes $\phi_{ab}$, $Q\in U(N^2)$ is a unitary transformation that diagonalizes $\Phi_{AB}$, and $X_{bB}=\d_{bB} x_b$ is a rectangular matrix whose only non-zero entries are the positive square roots of the eigenvalues of $\phi_{ab}$. Due to the invariance of such decomposition, the matrices $P$ and $Q$ are not unique, hence one should restrict the angular variables to $(P,Q)\in (U(N) \times U(N^2)/(U(N^2-N))\times U(1)^N)$.
Including the Jacobian of the transformation $(\vph,\vphb)\to(X,P,Q)$ \cite{Morris:1990cq,Anderson:1990nw,DiFrancesco:2002mvz} we rewrite the partition function \eqref{eq:Z-0d} as
\be \label{eq:Z-0d-eig}
Z(g) = \f{1}{\cN} \int \left( \prod_{i=1}^N dx_i\, x_i^{2(N^2-N)+1} e^{-N^2( x_i^2 - \f{g}{2} x_i^4)} \right) \D(x^2)^2 \, ,
\ee
where the angular variables have been factored out, and canceled with the normalization. The leftover normalization is in the factor $\cN$, equal to the integral in \eqref{eq:Z-0d-eig} evaluated at $g=0$. Lastly, $\D(x^2)=\prod_{1\leq i<j\leq N}(x_i^2-x_j^2)$ is the standard Vandermonde determinant for the eigenvalues of $\phi_{ab}$. The latter is subdominant in the large-$N$ limit, but the factor $x_i^{2(N^2-N)}$ is not, and must be taken into account. However, unlike the Vandermonde determinant, such term does not couple different eigenvalues, hence in the large-$N$ limit we have a simple saddle point equation in which the eigenvalues are mutually decoupled:
\be
x_i-g x_i^3 - \f{1}{x_i} = 0 \;\;\;\; \text{for}\;\;\; i=1,\ldots, N\,.
\ee
The solutions are
\be
x_\pm^2 = \f{1\pm \sqrt{1-4g}}{2g} \, ,
\ee
in agreement with \cite{Benedetti:2015ara},\footnote{The eigenvalues of the intermediate field $H_{ab}$ in \cite{Benedetti:2015ara} are $a_\pm=\sqrt{g} x_\pm^2$, because $H_{ab}$ is conjugate to $\sqrt{g}\phi_{ab}$.} where they were obtained by an intermediate field representation of the square of $\phi_{ab}$, as we did in Sec.~\ref{sec:S2}. 
For $g\leq 1/4$, the free energy at leading order is obtained by evaluating the effective action for the eigenvalues on the saddle point solution $x_i=x_-$, $\forall i$, which as shown in \cite{Benedetti:2015ara} is the only minimum of the action (and contrary to $x_+$, it is regular at $g=0$):
\be
-\f{1}{N^3} \ln(Z(g)) = x_-^2 - \f{g}{2} x_-^4 - \ln (x_-^2) - 1\;, 
\ee
where the minus one comes from the normalization factor $\cN$.
By explicit check, the result above coincides with the one in \cite{Benedetti:2015ara}.

Now, what happens if we introduce instead an intermediate field representation for the square of $\Phi_{AB}$?
The partition function \eqref{eq:Z-0d} rewrites
\be \label{eq:Z-0d-H}
\begin{split}
Z(g) &= \int [d\vph d\vphb dH] e^{- N^2 \left(\vphb_{aA}\vph_{aA} -\sqrt{g} \vph_{aA}\vphb_{aB} H_{AB} + \f12 H_{AB} H_{BA}  \right)}\\
& = \int [dH] e^{- \f{N^2}{2} \Tr(H^2)-N\Tr\ln(1- \sqrt{g}H)}
 \;.
\end{split}
\ee
Diagonalizing the matrix $H$, and counting the trace as a contribution of order $N^2$ (since the matrix $H$ is of size $N^2$), we conclude that the logarithmic term in the action is subleading, while the logarithm of the Vandermonde determinant originating in the change of variables is of $N^4$ as the Gaussian part of the action. On the other hand, the Gaussian integral on $H$ is normalized to one by construction, hence the free energy at leading order (LO) (order $N^4$) is correctly zero. In order to obtain the first non-trivial contribution to the free energy, we have to go to next-to-leading order (NLO) (order $N^3$).
Let us denote the eigenvalues of $H$ by $y_i$, $i=1\ldots N^2$. Writing explicitly the normalization as before, the partition function \eqref{eq:Z-0d-H} now writes
\be \label{eq:Z-0d-H-eig}
Z(g) = \f{1}{\cN'} \int \left(\prod_{i=1}^{N^2} dy_i\right) e^{- S_{\rm eff}(y)}\;,
\ee
with
\begin{align} \label{eq:Seff}
S_{\rm eff}(y) &\equiv N^4 S_0(y) + N^3 S_1(y)\;,\\
S_0(y)  &= \f{1}{2 N^2} \sum_{i=1}^{N^2}  y_i^2   -\f{2}{N^4} \sum_{1\leq i<j\leq N^2} \ln |y_i-y_j| \;, \\
S_1(y) &=  \f{1}{N^2} \sum_{i=1}^{N^2} \ln(1- \sqrt{g}y_i) \;.
\end{align}
Both $S_0(y)$ and $S_1(y)$ are of order one, and the order in $N$ has been made explicit in \eqref{eq:Seff}. Notice that the subleading term (of order $N^3$) is dominant over the one-loop correction, which is of order $N^2$. Therefore, we will not need to compute any loop corrections, and the NLO free energy is given by simply evaluating the subleading term of the action on the LO saddle point.
In fact, the saddle point $\bar{y}$ has an expansion (omitting the indices)
\be
\bar{y} = y_{{}_{\rm LO}} +\f1N y_{{}_{\rm NLO}} +O\Big(\f{1}{N^2}\Big)\;,
\ee
and the action then expands as
\be
S_{\rm eff}(\bar{y}) =  N^4 S_0(y_{{}_{\rm LO}}) + N^3\left(\f{\p S_0}{\p y}{\big|}_{{y=y_{{}_{\rm LO}}}} \cdot y_{{}_{\rm NLO}}+ S_1(y_{{}_{\rm LO}}) \right) +O(N^2)\;.
\ee
Since $\f{\p S_0}{\p y}{\big|}_{{y=y_{{}_{\rm LO}}}}  =0 $ by definition of $y_{{}_{\rm LO}} $, and since $\ln (\cN') = -N^4 S_0(y_{{}_{\rm LO}})+O(N^2)$, we obtain
\be
-\f{1}{N^3} \ln(Z(g)) =  S_1(y_{{}_{\rm LO}}) +O\Big(\f{1}{N}\Big)\;.
\ee
In order to complete the calculation, we need the explicit solution $y_{{}_{\rm LO}}$, which is given in terms of the famous Wigner semicircle law. 
This is given in terms of the density of eigenvalues $\r(y) = \f{1}{2\pi} \sqrt{4-y^2}$, from which we obtain
\be
S_1(y_{{}_{\rm LO}}) = \int_{-2}^{+2} dy \r(y) \ln(1- \sqrt{g}y) = x_-^2 - \f{g}{2} x_-^4 - \ln (x_-^2) - 1\;,
\ee
precisely the result we obtained before.

In conclusion, we have learned that it is possible to obtain the same results in three different ways: singular-value decomposition, intermediate field on the $U(N)$ sector, and intermediate field on the $U(N^2)$ sector. The first one is maybe the most enlightening for the zero dimensional case, but it becomes not viable in higher dimensions, where the angular variables do not decouple due to the derivative term (the action has a global invariance, not a local one). The last one is instead definitely the most difficult, as one needs to go to NLO and use the matrix model solution. For these reasons, in Sec.~\ref{sec:S2} we have followed the second option, which is of course also the standard one.

There remains to comment on the question of spontaneous breaking of the $U(N^2)$ symmetry. The fact that the saddle point of the intermediate field $H$ is given by the Wigner law would seem to indicate a complete breaking of the symmetry: the eigenvalues are spread along the interval $[-2,+2]$, hence the matrix is far from being proportional to the identity, as demanded by invariance under the symmetry group.
However, it is wrong to identify the saddle point solution of $H$ with its expecation value. In fact, in going from \eqref{eq:Z-0d-H} to \eqref{eq:Z-0d-H-eig} we have canceled the integral over the unitary group with the normalization, but if we have an insertion of a non-invariant observable, such as $\la H_{AB} \ra$, the integral does not drop. Furthermore, since the unitary matrices appearing in the change of variables from $H$ to its eigenvalues have no action term (because the action is invariant), we cannot do anything but perform the full integral.
We have:
\be
\begin{split}
\la H_{AB} \ra &= \f{1}{\cN'} \int \left(\prod_{i=1}^{N^2} dy_i\right) e^{- S_{\rm eff}(y)} Y_{CD} \int [dU] U_{AC}U^\dagger_{DB}\\
&= \f{\d_{AB}}{N^2} \f{1}{\cN'} \int \left(\prod_{i=1}^{N^2} dy_i\right) e^{- S_{\rm eff}(y)} \sum_i y_i\\
&= 0\;,
\end{split}
\ee
where we introduced the diagonal matrix of eigenvalues $Y_{AB}=y_A \d_{AB}$ and used the known formula for integration over the unitary group
\be
 \int [dU] U_{AC}U^\dagger_{DB} = \f{1}{N^2} \d_{AB} \d_{CD} \;.
\ee
Lastly, owing to the fact that the distribution of eigenvalues in the Wigner law is symmetric around the origin, $\Tr[Y]=0$ when evaluated on the saddle point solution. 

To conclude, we should remark that a similar phenomenon of symmetry restoration (\`a la Mermin-Wagner) will happen also in $d>0$. In higher dimensions, due to terms with derivatives in the action, the angular variables do not drop out of the action in general because we only have a global symmetry, not a local one. However, in the intermediate field representation the derivatives only appear in the higher-dimensional generalization of the logarithmic term in the action \eqref{eq:Z-0d-H}, which is subleading in this case. Hence, the same reasoning as in the zero dimensional case above applies also to the higher dimensional case.
In brief, the $U(N^2)$ part of the symmetry group does not undergo spontaneous symmetry breaking.

%%%%%%%%%%%%%%%%%%%%%%%%%%%%%
\section{Details on two and three traceless matrices}
\label{app:traceless_matrices}
%%%%%%%%%%%%%%%%%%%%%%%%%%%%%
\subsection{Two traceless matrices}
Imposing the intermediate field to have the form
\begin{equation}
M_1 = m_1(\mathbb{1}_N-2 \mathbb{P}_N), \quad M_2 = m_2(\mathbb{1}_N-2\mathbb{P}_N) , \quad M_3 = 0 \; ,
\end{equation}
the equations on motion have, for $m_1$, the form
\begin{equation}
\f{N^2}{2}\left(\f{1}{\lt_p} - \f{4}{\pi^2}\right)(\pm m_1) + \f{N^2}{\pi^2}\left[(\pm m_1 + m_2)^2\arctan\f{1}{\pm m_1 + m_2} + (\pm m_1 -m_2)^2\arctan\f{1}{\pm m_1 - m_2} \right] = 0\;,
\end{equation}
which combined with a similar one exchanging $m_1$ and $m_2$, lead to 
\begin{gather}
\Box\left(\f{1}{\lt_p} - \f{4}{\pi^2}\right) +  \f{4}{\pi^2}\Box^2\arctan\f{1}{\Box} = 0\;,\\
\bigtriangleup \left(\f{1}{\lt_p} - \f{4}{\pi^2}\right) + \f{4}{\pi^2} \bigtriangleup^2\arctan\f{1}{\bigtriangleup}= 0\;,\\
\Box \equiv m_1+m_2 \; , \quad \bigtriangleup \equiv m_1 - m_2\;.
\end{gather}
This implies either $\Box=0$ or $\bigtriangleup=0$, which after renaming gives $m_1 = m_2 = m$, or if neither is zero we have $\Box=\bigtriangleup$ forcing $m_1 = m_2 = 0$.
In the non-trivial case, the consequence for the effective potential is that we compare
\begin{gather}
V_1= \f{1}{4\lt_p}m^2 + \f{1}{3\pi^2}\k(m),\\
V_2= \f{1}{2}\left[\f{1}{4\lt_p}\Box^2 + \f{1}{3\pi^2}\k(\Box)\right],\\
\kappa(x) = 2x^3\arctan\frac{1}{x} - 2x^2 - \log(1+x^2), 
\end{gather}
with $m$ and $\Box$, obeying the same algebraic equation linking them to $\lt_p$. Now the dominance of the single over the double traceless matrices appears straightforwardly, as soon as they acquire negative values, that is for $\lt_p>\pi^2/4$.

\subsection{Three traceless matrices}
Imposing the intermediate field to have the form
\begin{equation}
M_1 = m_1(\mathbb{1}_N-2\mathbb{P}_N) , \quad M_2 = m_2(\mathbb{1}_N-2\mathbb{P}_N) , \quad M_3 = m_3(\mathbb{1}_N-2\mathbb{P}_N) \;,
\end{equation}
the equations of motion can be shaped into 
\begin{gather}
\left(\f{1}{\lt_p} - \f{4}{\pi^2}\right)\a + \f{3}{\pi^2}\alphab + \f{2}{\pi^2}\left(\betab + \gammab + \deltab \right) =0\;,\\
\left(\f{1}{\lt_p} - \f{4}{\pi^2}\right)\b + \f{3}{\pi^2}\betab + \f{2}{\pi^2}\left(\alphab + \gammab + \deltab \right) =0\;,\\
\left(\f{1}{\lt_p} - \f{4}{\pi^2}\right)\g + \f{3}{\pi^2}\gammab + \f{2}{\pi^2}\left( \alphab + \betab +\deltab \right) =0\;,\\
\left(\f{1}{\lt_p} - \f{4}{\pi^2}\right)\d + \f{3}{\pi^2}\deltab + \f{2}{\pi^2}\left( \alphab + \betab +\gammab \right) =0\;,\\
\xb \equiv x^2\arctan\f{1}{x}\;,\\
\a \equiv m_1 + m_2 + m_3\;,\quad \b \equiv m_1 - m_2 + m_3\;,\quad \g \equiv m_1 + m_2 - m_3\;,\quad \d \equiv -m_1 + m_2 + m_3\;.
\end{gather}
We then rearrange into
\begin{equation}
\left(\f{1}{\lt_p} - \f{4}{\pi^2}\right)\a + \f{1}{\pi^2}\a^2\arctan\f{1}{\a}=(\a \rightarrow \b) =(\a \rightarrow \g) = (\a \rightarrow \d) \overset{!}{=} K\;,
\end{equation}
for $K$ some constant. 

First, if we assume that $K$ vanishes, then the variables that are non-zero must obey the single-traceless equation of motion \eqref{eq: single-traceless-eom}. And two different options come: $\a=\b$ (or $\a=\g$ or $\a=\d$) letting us with the double-traceless case $(m_1 = m_2 = m ~ ;  m_3 = 0)$ (or permutations), or $\b=\g$ (or $\g=\d$ or $\b=\d$) giving $m_1 = m_2 = m_3 = m$.\footnote{Of course, we also have the combination of the two options which results to $m_1 = m_2 = m_3 = 0$.}  

Secondly, for non-zero $K$ and introducing $\lt_p = (1+\eps)\pi^2/4$, we found that non-trivial solutions (not resulting in $\a=\b=\g=\d$) existed only in the range $\eps \in [0,\eps_*]$ with $\eps_* \approx 0.37$. In this region, again by numerical exploration, we found two non-trivial solutions:
\begin{itemize}
\item $(\a = 3\b,~ \b = \g =\d) \implies m_1 = m_2 = m_3 = m$, with associated equation of motion and potential
\begin{gather}
4 \left(1 - \f{1}{1 + \eps}\right) - m \left(9 \arctan\f{1}{3 m} + \arctan\f{1}{m}\right) = 0 \;,\\
V_3 = \f{3 m^2}{\pi^2 (1 + \eps)} + \f{\k(3 m) + 3 \k(m)}{12 \pi^2}\; ;
\label{eq: triple-traceless1}
\end{gather}
\item $(\a = 0.08\b,~ \d = -1.92 \b,~ \b = \g) \implies -0.46 m_1 = m_2 = m_3$, with associated equation of motion and potential
\begin{gather}
4 \left(1 - \f{1}{1 + \eps}\right)0.92 + m \left(0.08^2 \arctan\f{1}{0.08 m} - \arctan\f{1}{m}\right) = 0\;,\\
V_3^\prime = \left(\f{0.92^2}{2} + 1\right) \f{m^2}{\pi^2 (1 + \eps)} + \f{\k(0.08 m) + 2 \k(m) + \k(-1.92 m)}{12 \pi^2}.
\label{eq: triple-traceless2}
\end{gather}
\end{itemize}

We studied numerically the differences $V_1-V_3$ and $V_1-V_3^\prime$, and found both to be negative, hence we conclude that the single-traceless solution is dominant.

%%%%%%%%%%%%%%%%%

%----- Bibliography ----------------------
%\bibliographystyle{JHEP-3}
%\bibliography{refs-3dTGN}
%---------------------------------------------

\providecommand{\href}[2]{#2}\begingroup\raggedright\endgroup

%%%%%%%%%%%%%%%%%
\end{document}